\documentclass[11pt,a4paper]{article}
\usepackage{t1enc}
\usepackage[latin1]{inputenc}
\usepackage[english]{babel}
\usepackage{natbib}
\usepackage{amssymb,stmaryrd}
\usepackage{graphicx}

\begin{document}
\thispagestyle{empty}

\title{Paving the way for transitions  --- \\ a case for Weyl geometry }
\author{Erhard Scholz\footnote{University of Wuppertal, Department C, Mathematics, and Interdisciplinary Centre for History and Philosophy of Science; \quad  scholz@math.uni-wuppertal.de}}

\date{June 21, 2015 }
\maketitle
%\vspace{2.5mm}

\begin{abstract}
This paper presents three aspects by which 
 the Weyl geometric generalization of Riemannian geometry, and of Einstein gravity,  sheds light on actual questions of physics and its philosophical reflection.  After  introducing  the theory's principles,  it  explains how Weyl geometric gravity  
 relates to Jordan-Brans-Dicke theory. We then  discuss the  link between gravity and  the electroweak sector of elementary particle physics, as it looks from the Weyl geometric perspective.  Weyl's hypothesis of a preferred scale gauge, setting Weyl scalar curvature to a constant, gets new support from the interplay of the gravitational scalar field and the  electroweak   one (the Higgs field). This has surprising consequences for cosmological models. In particular it leads to a static (Weyl geometric) spacetime  with ``inbuilt'' cosmological redshift. This may be used for putting central features of the present  cosmological model into a wider perspective. 
\end{abstract}

\tableofcontents

\subsection*{1. Introduction}
\addcontentsline{toc}{section}{1. Introduction}
When {\em Johann Friedrich Herbart}  discussed  the ``philosophical study''  of science he demanded that the sciences should  organize their specialized knowledge about {core concepts} ({\em Hauptbegriffe}). Philosophy should then strive 
%\begin{quote}
%\ldots 
``\ldots{\em to  pave the way for adequate transitions between the concepts }\ldots''
%\end{quote}
 in order to establish an  integrated system of knowledge.\footnote{``\ldots  und gilt uns [im philosophischen Studium, E.S.], dem 
gem\"a\ss{}, {\em alle Bem\"uhung, zwischen den Begriffen die geh\"origen Ueberg\"ange zu bahnen \ldots } '' \cite[275,  emphasis in original]{Herbart:1807}.} In this way philosophy and the specialized sciences were conceived as a common enterprise. Only together they would be able to generate a connected system of knowledge and contribute to  the ``many-sidedness of education''  Herbart had in  mind. 

This is not exactly what  is usually understood by ``metatheory''; but the concept of the  workshop which gave rise to this volume was to go beyond the consideration of working theories in themselves and to reflect on possible mutual connections between different spacetime theories, and perhaps beyond. This  task   comes quite close to what Herbart demanded from  `speculation' as he understood it. In this contribution  I want to use the chance offered by the  goal of the workshop  to discuss how  {\em Weyl geometry} may help to `pave the way for transitions' between certain segments of physical knowledge. We   deal here with connections between theories some of which came into existence long after the  invention of Weyl geometry and are far beyond  Weyl's original intentions during the years 1918 to 1923. 

%%%%%%%%%%%%%%%%%%%%%%(05.07.2013)
Mass generation of elementary particle fields is one of the topics. In general relativity mass serves as the  active and passive charge of the gravitational field;  high energy physics has made huge progress in analyzing the basic dynamical structures which determine the energy content, and thus the gravitational charge, of field constellations. The connection between high energy physics and gravity is still wide open for further research. Most experts expect the crucial link between the two fields to be situated close to the Planck scale, viz shortly after the big bang, with the Higgs ``mechanism'' indicating a phase transition in the early universe. This need not  be so. The Weyl geometric generalization of gravity considered here indicates a more structural connection between gravitation and the electroweak scalar field, independent of  cosmological time.  
The dilationally invariant Lagrangians of (special relativistic) standard model fields translate to scale invariant fields on curved spaces in an  (integrable) Weyl geometry. The latter offers a  well adapted arena for studying the transition between gravity and standard model fields.  Scalar fields play a crucial role on both sides, the question will be to what extent they are interrelated mathematically and physically.
%%%%%%%%%%%%%%%%%%%%%%%%%%%%%%%%%%%

Similar, although still more general, questions with regard to the transition from conformal structures to  gravity  theory have already been studied by Weyl. In his 1921 article on the relationship between conformal and projective differential geometry \cite{Weyl:projektiv_konform} he argued that  his new  geometry   establishes a peculiar bridge between the two basic geometrical structures underlying general relativity, conformal and projective. The first one was and still is the mathematical expression of the causal structure (light cones) and the second one represents the most abstract mathematization of inertial structure (free fall trajectories under abstraction from proper time parametrization).  Weyl  indicated a kind of  `transition'  to a fully metric  gravity theory into which other dynamical fields, in his case essentially the electromagnetic one, could be integrated.  He showed that a Weylian metric is uniquely determined if its conformal and its projective structures are known. In principle, such a metric can be determined by physically grounded structural observations without any readings of clocks or measurements with rods; i.e.,  Weyl geometry allows  to establish a connection between causal structure, free fall and metrical geometry in an impressingly basic  way.  

To make the present contribution essentially self-contained, we start with a short description of  Weyl geometry, already with physical meaningful interpretations in mind,  exemplified by the well-known work of Ehlers/Pir\-ani/Schild (section 2). 
In a  first transition we see how   
Jordan-Brans-Dicke (JBD) theory with its  scalar field, `non-minimally' coupled to gravity, fits neatly into a Weyl geometric framework  (section 3).  The different ``frames'' of JBD theory correspond to different choices of scale gauges of the Weylian approach. Usually this remains unnoticed in the literature,  although the basic structural ingredients of Weyl geometry are presupposed and dealt with in a non-explicit way. 

%%%%%%%%%%%%%%%%%%%%%%%%05.07.2013
The link is made explicit in a Weyl geometric version of  generalized Einstein theory with a  non-minimally coupled scalar field, due to Omote, Utiyama, Dirac e.a. (WOD gravity),   introduced in section 4. Strong reasons speak in favour of its integrable version (iWOD gravity) close to, but not identical with,  (pseudo-) Riemannian geometry. An intriguing parallel between the Higgs field of electroweak theory and the scalar field of iWOD gravity comes into  sight if one includes
 the gravitational coupling into the potential  of the scalar field. This suggests to consider a  common biquadratic potential for the two scalar fiels (section 5). In its minimum, the ground state of the scalar field specifies a (non-Riemannian) scale choice of the  Weyl geometry which establishes units for measuring mass, length, time etc, and gives rise to the vacuum expectation value and mass of the Higgs field.
 
  In his correspondence with Einstein on the physical acceptability of his generalized geometry Weyl conjectured, or postulated, an adaptation of atomic clocks to (Weylian) scalar curvature. In this way, according to Weyl, measuring devices would indicate a scaling in which (Weylian) scalar curvature becomes constant (Weyl gauge).
This conjecture is  supported, in a  surprising way, by evaluating the potential condition of the gravitational scalar field. If, moreover, the gravitational scalar field `communicates' with the electroweak Higgs field, clock adaptation to the ground state of the scalar field gets a field theoretic foundation in electroweak theory  (section 5.3, 5.4). The question is now open, whether such a transition between iWOD gravity and electroweak theory  indicates a  physical connection  or whether it is only an accidental  feature of the  two theories.

Reconsidering Weyl's scale gauge condition (constant Weylian scalar curvature) necessitates another look at cosmological models (section 6). The warping of Friedman-Robertson-Walker geometries can no longer immediately be interpreted  as an actual expansion of space (although that is not excluded). Cosmological redshift  becomes, at least partially, due to a field theoretic effect (Weylian scale connection). From such a point of view, much of the cosmological observational evidence, among it the cosmological microwave background and quasar distribution over redshift, ought to be reconsidered. The enlarged  perspective of integrable Weyl geometry and of iWOD gravity  elucidate, by contrast, how strongly
some  realistic claims of present precision cosmology are  dependent on  specific facets of the  geometrico-gravitational paradigm of Einstein-Riemann type. Many empirically sounding statements are insolvably intertwined with the data evaluation on this basis. Transition to a wider framework may be helpful to reflect these features  -- perhaps not only as a metatheoretical exercise (section 7).

%%%%%%%%%%%%%%%%%%%%%%%%%%%%%%%

\subsection*{2. On Weyl geometry and the analysis of EPS}
\addcontentsline{toc}{section}{2. On Weyl geometry and the analysis of EPS}
Weyl geometry is a generalization of Riemannian geometry, based on  two insights:  (i) The automorphisms of both, of Euclidean geometry  and  of special relativity, are the {\em similarities} (of Euclidean, or respectively of Lorentz signature) rather than the congruences. No unit of length is naturally given in Euclidean geometry, and likewise the basic structures of special relativiy (inertial motion and causal structure) can be given without the use of clocks and rods. (ii) The development of field theory and general relativity demands a conceptual implementation of this insight in a consequently {\em localized mode} (physics terminology).\footnote{In mathematical terminology, the implementation of a similarity structure happens at the {\em infinitesimal}, rather than at the local, level. For a concrete (``passive'') description of (i) and (ii) in a more physical language, see Dicke's postulate cited in section 3.1.}

Based on these insights, Weyl developed what he called {\em reine Infinitesimalgeometrie} (purely infinitesimal geometry) \cite{Weyl:InfGeo,Weyl:GuE}. Its basic ingredients are a conformal generalization of a  (pseudo-) Riemannian metric $g = (g_{\mu \nu}) $ by allowing point-dependent rescaling $\tilde{g}(x)= \Omega  (x)^2 \, g(x)$ with a nowherere vanishing (positive)  function $\Omega  $, and a scale (``length'') connection  given by a differential form $\varphi = \varphi_{\mu } dx^{\mu}$, which has to be gauge transformed $\tilde{\varphi} = {\varphi}-  d \log \Omega {} $ when rescaling $(g_{\mu \nu})$. The scale connection $(\varphi_{\mu})$ expresses how to compare lengths of vectors (or other metrical quantities) at two infinitesimally close points, both measured in terms of  a scale, i.e.,   a representative $(g_{\mu \nu})$ of the conformal class.\footnote{For more historical and philosophical details see, among others, \cite{Vizgin:UFT,Ryckman:Relativity,Scholz:Connections}, from the point of view of  physics \cite{Adler/Bazin/Schiffer,Blagojevic:Gravitation,%
Quiros:2013,Quiros:2014,Scholz:AdP}, and for the view of  differential geometers \cite{Folland:1970,Higa:1993,Gilkey_ea} (as a  short selection in all three categories).}

\subsubsection*{2.1 Scale connection, covariant derivative, curvature}
\addcontentsline{toc}{subsection}{2.1 Scale connection, covariant derivative, curvature}
Metrical quantities in Weyl geometry are directly comparable only if they are  measured at the same point $p$ of the manifold. Quantities measured at different points $p \neq q$ of finite, i.e., non-infinitesimal distance can be metrically compared only after an integration of the scale connection along a path from $p$ to $q$.  Weyl realized that this structure is compatible with  a uniquely determined affine connection $\Gamma = (\Gamma ^{\mu }_{\nu \lambda } )$ (the affine connection of Weylian geometry). If  ${}_g\Gamma^\mu _{\nu \lambda }$ denotes the Levi-Civita connection of the Riemannian  part $g$ only,  the Weylian affine connection is given by 
\begin{equation} \Gamma^{\mu }_{\nu \lambda } =   {}_g\Gamma^\mu _{\nu \lambda } + \delta ^{\mu }_{\nu } \varphi _{\lambda } +
\delta ^{\mu }_{\lambda } \varphi _{\nu } - g_{\nu \lambda } \varphi^{\mu }.\label{Levi-Civita}  
\end{equation} %\label{fn Levi-Civita} 
The {\em covariant derivative} with regard to $\Gamma$, will be denoted by $ \nabla = \nabla_{\Gamma }$. A change of scale  neither  changes the connection (the left hand side of (\ref{Levi-Civita})) nor the covariant derivative;  only the composition from the underlying Riemannian part and the corresponding scale connection (right hand side) is shifted.

Curvature concepts known from ``ordinary'' (Riemannian) differential geometry follow, as every connection defines a unique curvature tensor. The Riemann and Ricci  tensors, $Riem, Ric$, are  scale invariant by construction, although their expressions contain terms in $\varphi$. On the other hand,  the scalar curvature involves ``lifting'' of indices by the inverse metric and is thus scale covariant of weight $-2$ (see below). 

Field theory gets slightly more involved in Weyl geometry, because for vector and tensor fields (of ``dimensional'' quantities) the appropriate scaling behaviour under change of the metrical scale has  to be taken into account. If a field, expressed by $X$ (leaving out indices) with regard to the metrical scale $g(x)= (g_{\mu \nu}(x))$ transforms to
$ \tilde{X}= \Omega ^k X $  with regard to the scale choice $\tilde{g}(x)$ as above, $X$ is called a {\em scale covariant} field of {\em scale}, or {\em Weyl  weight} $w(X) := k$  (usually an integer or a fraction). Generally the covariant derivative, $\nabla X$,  of a scale covariant quantity $X$ is not scale covariant. However, scale covariance can be reobtained  by adding a weight dependent term. Then the {\em scale covariant derivative} $D$ of a scale covariant field $X$ is defined by
\begin{equation} DX := \nabla X + w(X)\varphi  \otimes X \; . \label{scale covariant derivative} \end{equation}
For example, $\nabla g$ is not scale covariant, but $Dg$ is. Moreover, one finds that $Dg=\nabla g + 2  \varphi \otimes g=0$; i.e., in Weyl geometry  {\em  $g$ appears} no longer {\em constant} with regard to the  derivative $\nabla $  but {\em with regard to the scale covariant derivative $D$}. 

In  physics literature an affine connection $\Gamma$ with $\nabla_{\hspace{-0.2em}\Gamma } \, g \neq 0$ is usually regarded as  ``non-metric'', and $\nabla_{\hspace{-0.2em}\Gamma }\, g$ is considered  its non-metricity.\footnote{See the contribution by F. Hehl, this volume.} 
These concepts hold in the Riemannian approach. In Weyl geometry, in contrast,   
\begin{equation} \nabla  g = - 2 \varphi \otimes g \quad  \Longleftrightarrow \quad D g = 0 \label{compatibility} \end{equation}  
expresses the {\em compatibility} of the affine connection $\Gamma$ with the {\em Weylian metric} represented by the pair $(g, \varphi)$.

{\em Geodesics}  can be  invariantly defined as autoparallels by  the Weyl geometric affine connection (so did Weyl himself). But but one can just as well,  in our context even better, consider scale covariant geodesics of weight $-1$ (see section 6.1). 

 Under a change of scale $g \mapsto \tilde{g} = \Omega^2 g$ and the accompanying gauge transformation for the scale connection $\varphi \mapsto \tilde{\varphi}=\varphi - d \log \Omega$, the compatibility condition transforms consistently, $\nabla_{\Gamma } \tilde{g} = - 2 \tilde{\varphi} \otimes \tilde{g}$. Equ. (\ref{compatibility}) ensures, in particular, that geodesics (i.e., auto-parallels) with initial direction along a nullcone of the conformal metric remain directed along the nullcones. This is the most important geometric feature of metric compatibility in Weyl geometry.\footnote{Weyl understood the  compatibility of the scale connection with the metric in the sense that   parallel transport of a vector $X(p)$ by the affine connection along a path $\gamma $ from $p$ to $q$ to $X(q)$ leads to consistency with length transfer along the same path. Compare the compatibility condition given, in a different mathematical framework, by \cite{EPS}. \label{fn compatibility} }

\subsubsection*{2.2 Weyl structures and integrable Weyl geometry (IWG)}
\addcontentsline{toc}{subsection}{2.2 Weyl structures and integrable Weyl geometry (IWG)}
In the more recent mathematical literature a {\em Weyl structure} on a manifold is defined by a  
 pair $(\mathcal{C}, \nabla)$ consisting of a {\em conformal structure} $\mathcal{C}=[g]$ (an equivalence class of pseudo-Riemannian metrics) and the covariant derivative of  a {\em torsion free linear connection} $\nabla$, constrained by the condition
\[ \nabla g + 2 \varphi_g \otimes g  = 0 \; ,\]
with a differential 1-form $\varphi_g$ depending on $g \in \mathcal{C}$.\footnote{ \cite{Higa:1993,Calderbank:2000,Ornea:2001,Gilkey_ea}} The change of the conformal representative
$g \mapsto \tilde{g}= \Omega ^2 \, g$ is accompanied by a change of the 1-form 
\begin{equation} \varphi_{\tilde{g}}= {\varphi_{g}}-  d \log \Omega {}\; ,  \end{equation}
 i.e., by a ``gauge transformation'' as introduced  by Weyl  in \cite{Weyl:GuE}. Formally, a {\em Weyl metric} consists of an equivalence class of pairs $(g, \varphi_g)$ with scale and gauge transformations defining the eqivalences.  Given the scale choice $g \in \mathcal{C}$,  $\varphi_g$ represents the scale connection,. 

In Weyl's view of a strictly ``localized'' (better: infinitesimalized) metric, metrical quantities at different points $p$ and $q$ can be compared only by a   ``transport of lengths standards'' along a path $\gamma $ from $p$ to $q$, i.e., by   multiplication  with a factor 
 \begin{equation}  l(\gamma )  = e^{\int_0^1 \varphi (\gamma ') } \, . \label{scale transfer} \end{equation} 
$l(\gamma ) $ will be called the {\em length} or {\em scale transfer}  function (depending on $p, q$ and $\gamma$).
The {\em curvature} of the {\em scale connection}  is simply  the exterior differential, $f =d \varphi$ with components, $f_{\mu \nu}= \partial _{\mu} \varphi_{\nu}- \partial _{\nu} \varphi_{\mu}$, where $ \partial _{\mu}  := \frac{ \partial  }{ \partial {x^\mu} }  $.

For vanishing scale curvature, $f=0$, the scale transfer function can be integrated away, i.e., there exist local choices of the scale, $\tilde{g}$, with vanishing scale connection, $\varphi_{\tilde{g}}=0$.  In this case one deals with {\em integrable Weyl geometry} (IWG). Then the Weyl metric may be locally represented  by a Riemannian metric;\footnote{Here ``local'' is used in the sense of differential geometry, i.e., in (finite) neighbourhoods. Physicists usage of ``local'', in contrast, refers in most cases to point-dependence or ``infinitesimal''  neighbourhoods. In the following, both language codes are used, not always with further specification. The respective meaning will be clear from the context. }
 we call this the {\em Riemann gauge} (equivalently {\em Riemannian scale choice}) of an integrable Weyl metric. In this gauge the Weylian curvature tensor does not contain terms in $\varphi$.
 For integrable Weyl geometry vanishing of the Riemann tensor, $Riem =0$ is of course equivalent to local flatness.
 
Whether  a reduction to Riemannian geometry makes sense physically, depends on the field theoretic content of the theory.  If a scalar field plays a part  in determining the scale --- physically speaking, if scale symmetry is ``spontaneously'' broken by a scale covariant scalar field ---   the result  may  well be  different from Riemannian geometry  (see below, sections 4ff.). 

\subsubsection*{2.3 From Ehlers/Pirani/Schild to Audretsch/G\"ahler/Straumann}
 \addcontentsline{toc}{subsection}{2.3 From Ehlers/Pirani/Schild to Audretsch/G\"ahler/Straumann}
 Weyl  originally hoped to  represent the potential of the electromagnetic field by a scale connection and to achieve a geometrical unification of gravity and electromagnetism by his ``purely infinitesimal'' geometry. The physical difficulties of this appoach, usually presented as  outright inconsistencies with observational evidence, have been discussed in the literature \cite{Vizgin:UFT,Goenner:UFT}. But, of course, there is no need to bind the usage of Weyl geometry  to this specific, and outdated, interpretation. Since the early 1970s a whole, although  minoritarian and heterogeneous, literature of Weyl geometric investigations in the foundations  of gravity  has emerged. In this contribution I want to take  up, and pursue a little further, an approach going back to M. Omote, R. Utiyama, and P.A.M. Dirac, which was later extended in different directions (section 4, below).\footnote{The interpretation of the quantum potential in Weyl geometric terms proposed by  \cite{Santamato:WeylSpace,Santamato:KG} and others indicate  a completely different route of attempted ``transitions'' than reviewed here. It is not further considered in the following.}
 But before we follow these more specific lines we have to briefly review  the foundational aspects of Weyl geometry for gravity theory analyzed in the seminal paper of J. Ehlers, F. Pirani and A. Schild (1972) (EPS). 
 
 Like Weyl in 1921, these three authors based their investigation on the insight that the causal structure of general relativity is mathematically characterized by a  conformal (cone) structure, and the  inertial structure of point particles by a  projective path structure. They investigated the interrelation of the two structures from a foundational point of view in a methodology sometimes called a ``constructive axiomatic'' approach. Their axioms  postulated  rather general properties for  these two structures and demanded their 
compatibility.  EPS concluded that these properties suffice for specifying a unique Weylian metric \cite{EPS}.\footnote{For the compatibility see fn. \ref{fn compatibility}. A recent  commentary of the paper is given in \cite{Trautman:EPS}. How $f(R)$ theories of gravity may lead back to the EPS paper is discussed in \cite{Capoziello_ea:EPSetc}.}
 The axioms of Ehlers, Pirani and Schild were motivated by the physical intuition of inertial paths (of classical particles) and the causal structure. Other authors investigated connections to quantum physics. J. Audretsch, F. G\"ahler, N. Straumann (AGS) found that wave functions (Klein-Gordon and Dirac fields) on  a Weylian manifold behave acceptable only in the integrable case. As a criterion of acceptability they studied the streamlines of wavefront developments in an WKB approximation (WKB: Wentzel-Kramers-Brioullin)  and found that, for $\hbar \rightarrow 0$, the streamlines converge to geodesics if and only if $d\varphi = 0$, i.e., in the  case of an {\em integrable } Weyl metric \cite{AGS}. Therefore the integrability of the Weyl structure  seems necessary for consistency between the geodesic principle of classical particles and the decoherence view of the quantum to classical transition.

The gap between the structural result of EPS (Weyl geometry in general) and the 
pseudo-Riemannian structure of ordinary (Einstein) relativity was  considerably reduced in the sense of integrability, but still it was not clear that the Riemannian scale choice of IWG had to be chosen. The selection of Riemannian geometry remained ad hoc and was  not based on deeper insights. It  had to be stipulated by an additional postulate involving clocks and rods. The transition from the EPS axiomatics  to Einstein gravity still contained a methodological jump and relied on reference to observational instruments external to the theory, which Weyl wanted to exclude from the foundations of general relativity.\footnote{Although in his  1918 debate with Weyl, Einstein insisted on the necessity of clock and rod measurements in general relativity as the empirical basis for the physical metric,  he admitted that rods and clocks should not be accepted as fundamental. He reiterated this view until late in his life \cite[555f.]{Einstein/Schilpp}, cf. \cite{Lehmkuhl:2014}.}
So even after the work of EPS and their successors the question remained whether the transition to Riemannian geometry and Einstein gravity is the only one possible. Alternatives were sought for by a different group of authors who started more or less simultaneous to  EPS,  investigating alternatives based on a scale invariant Lagrangian (section 4) similar to the one studied by Jordan, Brans, and Dicke in the Riemannian context. It was not noticed at the time that even the latter can  be  analyzed quite naturally  in the framework of Weyl geometry. 

 \subsection*{3. Jordan-Brans-Dicke theory in Weyl geometric perspective}
 \addcontentsline{toc}{section}{3. Jordan-Brans-Dicke theory in Weyl geometric perspective}
In the early 1950s and 1960s P. Jordan,  later R. Dicke and C. Brans (JBD) proposed a widely discussed   modification of Einstein gravity.\footnote{\cite{Jordan:Schwerkraft,Brans/Dicke,Dicke:1962}; for surveys on the actual state of JBD theory and its applications to cosmology see \cite{Fujii/Maeda,Faraoni:2004}, for a participant's recollection of its history \cite{Brans:Roots}.}
 Essential for their approach was a (real valued) scalar field $\chi $, coupled to the traditional Hilbert action  with Lagrangian density 
 \begin{equation} \mathcal{L_{JBD}} = (\chi  R - \frac{\omega }{\chi }\partial ^{\mu}\chi \, \partial _{\mu}\chi  ) \sqrt{ | det \,  g|}  \; , \label{L_JBD}
\end{equation} 
where $\omega $ is a free parameter of the theory. For $\omega \rightarrow \infty $ the theory has Einstein gravity as limiting case. 
All three authors allowed for conformal transformations, $\tilde{g} = \Omega ^2 g$, under which their scalar field $\chi $ transformed with weight $-2$ (matter fields and energy tensors $T$ of weight $w(T)=-2$ etc.).\footnote{Weights rewritten in adaptation to our convention.} 
Jordan took up the discussion of conformal transformations  only in the second edition of his book \cite{Jordan:Schwerkraft},  after Pauli had made him aware of such a possibility. Pauli knew Weyl geometry very well, he was one of its experts  already as early as  1919 but neither he nor Jordan or the US-American authors  looked at JBD theory from that point of view. 

\subsubsection*{3.1 Conformal rescaling in JBD theory}
\addcontentsline{toc}{subsection}{3.1 Conformal rescaling in JBD theory}
For introducing  conformal rescaling Dicke argued  as follows:
\begin{quote}
It is evident that the particular values of the units of mass, length, and time employed are arbitrary and that  the {\em  laws of physics must be invariant under a general coordinate dependent  change of units}  \cite[2163]{Dicke:1962}[emph. ES].
\end{quote}
By ``coordinate dependent change of units'' Dicke indicated a point dependent rescaling of basic units. In the light of 
 the relations  established by the fundamental constants (velocity of light $c$, (reduced) Planck constant $\hbar$, elementary charge $e$ and Boltzmann constant $k$) all units can be expressed in terms of one independent fundamental unit, e.g. time, and the fundamental constants (which, in principle can be given any constant numerical value, which then fixes the system).\footnote{The present  revision of the  international standard system SI is heading toward implementing  measurement definitions with time as only fundamental unit, $u_T = 1\, s$ such that ``the ground state hyperfine splitting frequency of the caesium 133 atom  $\Delta\nu(^{133}Cs )_{\mbox{hfs}}$ is exactly $9\,192\,631\,770$ hertz''  \cite[24f.]{SI:2011}. In the ``New SI'', four of the SI base units, namely the kilogram, the ampere, the kelvin and the mole, will be redefined in terms of invariants of nature; the new definitions will be based on fixed numerical values of the Planck constant, the elementary charge, the Boltzmann constant, and the Avogadro constant (www.bipm.org/en/si/new$_{-}$si/). The redefinition of the meter in terms of the basic time unit by means of the fundamental constant $c$ was implemented already in 1983. Point dependence of the time unit because of locally varying gravitational potential will be inbuilt in this system. For practical purposes it can be outlevelled by reference to the {\em SI  second on the geoid} (standardized by the International Earth Rotation and Reference Systems Service IERS).  \label{fn SI} } 
 Thus only one essential scaling degree of units remains and 
Dicke's  principle of an arbitrary, point dependent unit choice came down to a ``passive'' formulation of Weyl's  localized similarities  in the framework of his scale gauge geometry.\footnote{Compare principles (i) and (ii) at the beginning of section 2.}
 It was not so clear, however,  how Dicke's postulate that the ``laws of physics
must be invariant'' under point dependent rescaling ought to be understood in JBD theory. Its modified Hilbert term  was, and is, not  scale invariant and assumes correction terms under conformal rescaling (vanishing only for $\omega=-\frac{3}{2}$).

On the other hand, the principles of JBD gravity were moved even closer to  Weyl geometry  by
 all three proponents  of this approach considering it as self-evident that the 
{\em  Levi-Civita connection} $ \Gamma := \, _g\hspace{-0.05em}\Gamma $ of the Riemannian metric $g$ in
 (\ref{L_JBD}) remains {\em unchanged}  under conformal transformation of the metric. Probably the protagonists considered that as a natural outcome of   assuming  invariance of the ``laws of nature''  under conformal rescaling.\footnote{If the  trajectories of bodies are governed by the  gravito-inertial ``laws of physics'' they should not be subject to change under transformation of units. The same should hold for the  affine connection  which can be considered a mathematical concentrate of these laws.} 
 In any case, they kept the affine connection  $ \Gamma$   fixed    and rewrote it in terms of  the Levi-Civita connection  $_{\tilde{g}}\hspace{-0.05em}\Gamma $ of the rescaled metric, $\tilde{g}=\Omega^2 g$, with additional terms  in partial derivatives of $\Omega $. Let us summarily denote these additional terms by  by $\Delta (\partial  \Omega)$,\footnote{For our purpose the explicit form of $\Delta (\partial  \Omega)$ is not important. R. Penrose noticed that the  additional terms of the (Riemannian) scalar curvature are exactly cancelled by the partial derivative terms of the kinematical term of $\chi$ if and only if $\omega = - \frac{3}{2}$. In this case the Lagrangian (\ref{L_JBD}) is conformally invariant \cite{Penrose:1965}.}
 then
\[ \Gamma = \, _{\tilde{g}}\hspace{-0.05em}\Gamma  + \Delta (\partial  \Omega) \; . \] 
  Conformal rescaling, in addition to a fixed affine connection, have become {\em basic tools of  JBD theory}.

\subsubsection*{3.1 IWG as implicit framework of JBD gravity}
 \addcontentsline{toc}{subsection}{3.1 IWG as implicit framework of JBD gravity}
The variational principle (\ref{L_JBD}) of JBD gravity determines a connection with covariant derivative $\nabla= \, _{g}\hspace{-0.05em}\nabla$ and a scalar  field $\chi$. The theory allows for conformal rescalings of $g$ and $\chi$ without changing $\nabla$. That is, JBD theory {\em specifies a Weyl structure} $(\mathcal{C},\nabla)$ with $\mathcal{C}=[g]$. Transformation between  different frames happen in this framework, even though this  remains unreflected  by most of its authors.  

In the JBD tradition, a choice of units is  called a {\em frame}. In terms of Weyl geometry such a frame corresponds to the selection of a scale gauge. Two frames play a major role:
\begin{itemize}
\item {\em Jordan frame}: the one in which $\nabla = \, _g\hspace{-0.2em}\nabla $ \quad ($g$ the metric of (\ref{L_JBD})), i.e.,  the affine connection is the Levi-Civita one of the Riemannian metric,
\item {\em Einstein frame}: the one in which $\tilde{\chi }= const$; then the affine connection is different from the Levi-Civita one of the reference metric. 
\end{itemize}
 The Jordan frame  is  such that, by definition, the dynamical affine connection is identical to  the  Levi-Civita connection of  $g$. 
Expressed in Weyl geometric terms,  this implies  vanishing of the scale connection,    $\varphi = 0$. Thus this frame correponds to what we have called the {\em Riemann gauge} of the underlying integrable Weylian metric (section 2). In Einstein frame  the scalar field ($\neq$ 0 everywhere) is scaled to a constant; we may call this the {\em scalar field gauge}. Another terminology for it is {\em Einstein gauge}. In this gauge, the gravitational ``constant'' appears as a true constant, contrary to Jordan's motivation.  By obvious reasons, Jordan  tended to prefer the  other frame; thus its name. 

Clearly in the Einstein frame JBD gravity does not reduce to Einstein gravity, as the affine connection is deformed  with regard to the metrical  component of the gauge. 
Scalar curvature in Einstein frame can easily be expressed in terms of Weyl geometrical quantities, but usually it is not. Practitioners of JBD theory  prefer to write everything in  terms of $\tilde{g}$,  take its Levi-Civita connection $_{\tilde{g}}\hspace{-0.01em}\Gamma$ as representative for the gravito-inertial field and consider the modification terms as arising from the transformation from Riemann gauge  to scalar field gauge. Sometimes they appear as additional (``fifth'') force.\footnote{For a critical discussion see \cite{Quiros_ea:2012}.}

From our point of view, we observe:
\begin{itemize}
\item Structurally, JBD theory presupposes and works in an {\em integrable Weyl structure}, although its practitioners usually do not notice.\footnote{A discussion from a slightly different view can  be found  in \cite{Romero_ea:2011,Quiros_ea:2012,Almeida/Pucheu:2014}.}
\item {\em Scale  covariance}, not scale invariance, is often the game of JBD theoreticians. That lead to a debate (sometimes confused), which frame should be considered as ``physical'' and which not. Jordan frame used to be the preferred one.  In the recent literature of JBD some, maybe  most,   authors argue in favor of  Einstein frame as ``physical'' \cite{Faraoni:Frames}.
\item Some authors studied the  conformally invariant version of the JBD Lagrangian, corresponding to $\omega =-\frac{3}{2}$, and investigated the hypothesis of a conformally invariant theory of gravity at high energies, which gets ``spontaneously broken'' by the scalar field taking on a specific value \cite{Deser:1970,Englert/Gunzig:1975}. That was achieved by  adding additional polynomial terms in $\chi$ with coefficients usually of ``cosmological'' order of magnitude. Problems arose in the conformal JBD approach from the sign of  $\omega $; a negative sign indicated a ``ghost field'' with negative energy \cite[5]{Fujii/Maeda}.\footnote{Some authors  choose to switch  the sign of the ``gravitational constant'', e.g. \cite[250]{Deser:1970}. This strategy indicated  that there is  a basic problem for the conformal JBD approach ($\omega =-\frac{3}{2} $) in spite of its attractive basic idea. \label{fn wrong sign}}
\item Empirical high precision tests of gravity in the solar system concentrated on the Jordan frame  and found increasingly high bounds for the  parameter $\omega$. To the disillusionment  of JBD practitioners, $\omega$   was found to be $  > 3.6 \cdot 10^3$  at the turn of the millenium \cite{Will:LivingReviews}; today these values are even higher. So the leeway  for JBD theory {\em in Jordan frame}  deviating from Einstein gravity became increasingly reduced. That does not hinder  authors in cosmology to assume  Jordan frame models for the expansion of universe shortly after the big bang.\footnote{E.g. \cite{Guth/Kaiser,Kaiser:1994,Bezrukov/Shaposhnikov:2007,Kaiser:2010}.}
Shortly after the big bang, the world of mainstream cosmology seems to be Feyerabendian.
\end{itemize}
From the Weyl geometric perspective,  a criterion of scale invariance for observable quantities  supports preference of the Einstein frame. In any case,  Weyl geometry is a conceptually better adapted  framework  for JBD gravity than Riemannian geometry. Perhaps that was felt by some physicists at the time. Be that as it may,  about a decade after the rise of JBD  theory  two groups of   authors in Japan and in Europe, indepently of each other, started to study a similar type of coupling between scalar field and gravity in a  Weyl geometric  theory of gravitation. 

\subsection*{4. Weyl-Omote-Dirac gravity and its integrable version (iWOD)}
\addcontentsline{toc}{section}{4. Weyl-Omote-Dirac gravity and its integrable version (iWOD)}
In 1971 M. Omote proposed a Lagrangian field theory of gravity with a scale covariant scalar field  coupling to the Hilbert term like in JBD theory, but now explicitly formulated in  the framework of Weyl geometry. A little later R. Utiyama and others took up the approach for investigations aiming at an overarching theory of strongly interacting fields and 
gravity.\footnote{\cite{Omote:1971,Omote:1974,Utiyama:1975I,Utiyama:1975II,Hayashi/Kugo:Weyl_field} --- thanks to F. Hehl to whom I owe the hint to Omote's works.}
Indepently P.A.M. Dirac initiated a similar line of  research with a look at possible connections between fields of high energy  physics, gravity  and cosmology \cite{Dirac:1973}. It did not take long until the idea of a spontaneously broken conformal gauge theory of gravitation was also considered in the framework of Weyl 
geometry and brought into first contact with the rising standard model of elementary particle physics \cite{Smolin:1979,Nieh:1982,HungCheng:1988,Hehl_ea:Progress1989}. Important for this move seemed to be that the obstacle of a negative energy (``ghost'') scalar field or wrong sign of the gravitational constant,  arising in  the strictly conformal version of JBD theory, could be avoided in this framework.\footnote{Cf. fn. \ref{fn wrong sign}.}  
Here we are not interested in historical details, but aim at sketching the  potential of the approach from a more or less philosophical point of view.\footnote{For a first rough outline of the history see   \cite{Scholz:Mainz}. For a commented source collection of much wider scope  \cite{Blagojevic/Hehl}.}

\subsubsection*{4.1 The Lagrangian of WOD gravity}
 \addcontentsline{toc}{subsection}{4.1 The Lagrangian of WOD gravity}
The affine connection of  Weyl geometry is scale invariant; the same holds for its Riemannian curvature $Riem = (R_{\lambda \mu \nu }^{\kappa})$ and  the Ricci tensor $Ric = (R_{ \mu \nu })$ as its contraction.\footnote{We  use abbreviated symbols of geometrical objects, $Riem, Ric, \varphi, \nabla$ etc. together with their indexed coordinate description. The whole collection of indexed quantities will be denoted by round brackets like in matrix notation, e.g. $Ric = (R_{ \mu \nu })$ or $\varphi= (\varphi_1, \ldots , \varphi_n)$, in short $\varphi=(\varphi_{\mu})$. The latter is somehow analogous to $\varphi_{\mu}$ in ``abstract index notation'', often to be found in the literature.  In our notation the bracketed symbol stands for the whole collection of indexed quantities,  the unbracketed symbol for a  single indexed quantity $\varphi_{\mu} \in \{\varphi_1, \ldots, \varphi_n \}$. }
  Scalar curvature $R=g^{\mu \nu}R_{\mu \nu}$ is  scale covariant of weight $w(R)= w(g^{\mu \nu}) = -2$. Coupling of  a norm squared real or complex scalar 
field\footnote{Later the scalar field is allowed to take values in an isospin $\frac{1}{2}$ representation of the electroweak group, section 4.5. } 
$\phi$ of weight $-1$ to the scalar curvature of Weyl geometry  gives, for the  Lagrangian density of the modified Hilbert term
\begin{equation} \mathcal{L}_{HW}= {L}_{HW} \sqrt{ | det \,  g|} = -  \frac{1}{2}\xi^2  |\phi|^2 R \sqrt{ | det \,  g|}\, ,
\end{equation} 
 a total weight  $-2-2+4=0$ and thus scale invariance.\footnote{$w( \sqrt{ | det \,  g|})= \frac{1}{2}\,4\cdot 2 =4$, $w({L}_{HW})= - 2 - 2=-4$}
If  $R$ denotes just that of  Riemannian  geometry and if one adds the kinematical term of the scalar field, Penrose's criterion for  conformal invariance only holds  for $\alpha = - \frac{1}{6}$. It is crucial to realize that in the Weyl geometric framework local scale invariance holds for {\em any coefficient}. 

Conformal rescaling  leads to different ways of decomposing covariant or invariant terms into  contributions from the Riemannian component $g$ and the scale connection $\varphi$ of a representative (a ``scale gauge'') $(g,\varphi)$ of the Weylian metric. We characterize these components by subscripts put in front; e.g.  for scalar curvature the  decomposition is  summarily written  as $ R = _g\hspace{-0.3em}R+ _{\varphi}\hspace{-0.3em}R $, with $_g\hspace{-0.2em}R$ the scalar curvature of the Riemannian part $g$ of the metric alone and $_{\varphi}\hspace{-0.2em}R$ the term due to the respective scale connection. For dimension $n=4$ of spacetime one obtains (independently of the signature)
 \begin{equation} _{\varphi}\hspace{-0.1em}R  = - (n-1)(n-2) \varphi_{\lambda } \varphi^{\lambda }  - 2(n-1) _g \hspace{-0.2em}\nabla _{\lambda }\varphi^{\lambda } 
= - 6 \varphi_{\lambda } \varphi^{\lambda }  - 6 _g \hspace{-0.2em}\nabla _{\lambda }\varphi^{\lambda } \; ,
 \label{scalar curvature} \end{equation}
where  $_g \hspace{-0.2em}\nabla $  denotes the covariant derivative (Levi-Civita connection)  of the Riemannian part $g$ of the metric. %\label{fn R}
Of course, the merging of scale dependent terms to scale invariant aggregates is of primary  conceptual import, besides being calculationally advantageous.\footnote{The authors of the 1970s usually did not use the aggregate notation.} 

The gradient term of the scalar field in Omote-Dirac gravity is modelled after the kinematical term of a  Klein-Gordon field:
\begin{equation}  L_{\phi} =   \epsilon _{sig}\frac{1}{2} D_{\nu } \phi^{\ast} D^{\nu } \phi    \, , \qquad   \mathcal{L_{\phi}} = {L_{\phi}} \sqrt{ | det \,  g|}
\end{equation} 
with scale covariant derivative $D_{\nu }\phi = (\partial _{\nu} - \varphi_{\nu}) \phi $, according to equ. (\ref{scale covariant derivative}), is  scale invariant, as $w(L_{\phi}) =-4$. Here $ \epsilon _{sig}$ specifies a signature dependent sign: $\epsilon _{sig} =  1$ for $sig = (1,3)$ i.e.,  $ { (+ - - -)}$  and  $\epsilon _{sig} =  -1$ for $sig = (3,1) \sim  (- + + + )$. In this paper we shall work with this kinematical term. In other contexts, e.g. in a Weyl geometric adaptation of the AQUAL approach to relativistic MOND dynmaics, one has to allow for other forms of $L_{\phi}$. In particular  a cubic gradient terms may lead to new insights in modified gravity on the galactic and the cluster level.\footnote{AQUAL stands for the ``aquadratic Lagrangian'' approach, which was the first attempt at a relativistic version of MOND dynamics \cite{Bekenstein/Milgrom:1984}. 
An adaptation to Weyl geomtric gravity is investigated in \cite{Scholz:MOND-like}. }

A polynomial potential for the scalar field $V(\phi)$ leads to a scale invariant Lagrange term if and only if the degree of $V$ is four, i.e., for a quartic monomial
\begin{equation} L_V = - \frac{\lambda}{4} |\phi|^4   \, , \qquad   \mathcal{L}_{V} = {L_{V}} \sqrt{ | det \,  g|}\; .
\end{equation}

Considering the scale connection $\varphi$ as a dynamical field, the ``Weyl field'' with its  quantum excitation, called ``Weyl boson'' or even ``Weylon'' by \cite{HungCheng:1988},  demands to add a Yang-Mills action for the scale curvature $f=(f_{\mu \nu})$:
\begin{equation} L_{_{YM} \varphi} =  - \frac{\beta }{4}f_{\mu \nu } f^{\mu \nu }   
\end{equation} 
 So did Omote, Dirac and later authors.\footnote{Dirac, curiously, continued  even in the 1970s to stick to the interpretation of the scale connection as  electromagnetic potential. No wonder that this prposal was not accepted even in the selective reception of his work.} 
 
The whole scale invariant Lagrangian of Weyl-Omote-Dirac gravity including the scalar field, neglecting for the moment further couplings to matter and interactions fields,  is given by 
\[ L_{WOD}=L_{R^2}+ L_{HW}+L_{\phi}+ L_V  + L_{YM} + L_{m}\, , \]
where $L_{R^2}$ contains all second order curvature contributions.  They seem to be necessary if one wants to study (perturbative) quantization, starting from this classical template.  $L_{m}$ denotes matter and interaction terms, for example the adapted  standard model fields, $L_m= L_{SM}$ (lifted to curved Weyl space).\footnote{Signs are chosen sucht that $\phi$ has positive energy density (no ghost field) \cite[5]{Fujii/Maeda}. In \cite[equ.(8.5)]{Blagojevic/Hehl} the coefficient  $\alpha $ has to be assumed negative -- compare with their source paper 8.3 (Nieh 1982), eqs. (2) and (7). For the role of $L_{R^2}$ in quantum gravity see \cite[18ff., 62ff.]{Capozziello/Faraoni} and, historically, \cite{Schimming/Schmidt}. For steps toward adapting the  standard model Lagrangian to Weyl geometry (basically by writing it locally scale invariant) see, among others, \cite{Drechsler/Tann,Nishino/Rajpoot:2004,Meissner/Nicolai,Quiros:2014,Bars/Steinhardt/Turok:2014}. } 
\begin{equation} L_{WOD}= L_{R^2} - \epsilon _{sig}\frac{1}{2}\xi^2 |\phi|^2 R - \frac{\lambda }{4}|\phi|^4 +  \epsilon _{sig}\frac{1}{2} D_{\nu } \phi^{\ast} D^{\nu } \phi     - \frac{\beta }{4}f_{\mu \nu } f^{\mu \nu } + 
L_m  \;  
\end{equation} 
Formally it contains  a Brans-Dicke like modified Hilbert action,  a ``cosmological'' term, quartic in $\phi$, and dynamical terms for the scalar field and the scale connection. The Weyl geometric expressions for scalar curvature and scale covariant derivative ensure scale invariance of the Lagrangian density  $\mathcal{L}_{WOD}=  L_{WOD}\sqrt{ | det \,  g|}$.
 Scale invariance forces the polynomial part of the potential with constant coefficients to be exclusively quartic.
Later we shall see that the assimilation of the standard model Lagrangian $L_{SM}$ to gravity makes it necessary to modify the potential term $L_V = V(\phi) = - \frac{\lambda}{4}$ by introducing a combined quartic potential $V(\phi, \Phi)$ for the gravitational scalar field and the Higgs field $\Phi$.

 \subsubsection*{4.2 From WOD to iWOD gravity}
 \addcontentsline{toc}{subsection}{4.2 From WOD to iWOD gravity}
 A closer look at  the WOD-Hilbert term shows that, because of equ. (\ref{scalar curvature}),   it contains a 
 mass-like term for the scale connection (the ``Weyl field''): 
\begin{equation}  \frac{1}{2}  m_{\varphi}^2 \, \varphi_{\lambda } \varphi^{\lambda } =  \frac{1}{2}  6 \xi^2 |\phi|^2 \varphi_{\lambda } \varphi^{\lambda } 
\end{equation}  

If WOD  describes a realistic modification of Einstein gravity, its  Hilbert term has to approximate the latter very 
well under the limiting  conditions $|\phi|\rightarrow const, \; \varphi \rightarrow 0$.
Then $\xi^2 |\phi|^2$ must be comparable to the inverse of the gravitational constant $\xi^2 |\phi|^2 \approx [\hbar c](8 \pi G)^{-1} = \frac{m_{pl}^2}{8 \pi} = M_{pl}^2$ with reduced Planck mass $M_{pl}$.\footnote{$m_{pl}^2= \frac{\hbar c}{G}$, with ``reduced'' $M_{pl}:=\sqrt{\frac{\hbar c}{8 \pi G}} $.}
Then   the ``Weylon'' (Cheng, Nishino/Rajpoot e.a.) turns out to be  sitting a little above the reduced Planck mass (but below the unreduced one):
\begin{equation} m_{\varphi} \approx  2.5 M_{pl} \approx 0.5\, m_{pl}
\label{mass Weylon} \end{equation} 
Variation of the Lagrangian shows that it satisfies a Proca equation with this tremendously high mass \cite{Smolin:1979,HungCheng:1988}. Because of the scaling behaviour of $\phi$ the Proca-like mass term does not destroy scale invariance of the Lagrangian.\footnote{Therefore the, otherwise interesting, discussion of the gravitational scalar field as a kind of ``St\"uckelberg compensator''by \cite{Nishino/Rajpoot:2009} seems a bit artificial.} 

If one assumes a physical role for the Weyl field, its (immediate) range, in the sense of its Compton wave length,  would  be restricted to  Planck scale physics. On all scales accessible to experiments and to direct observation  the {\em curvature of the Weyl field  vanishes effectively}. This result agrees  with the integrability result  of Audretsch, G\"ahler and Straumann on the compatibility of Weyl geometry with quasi-classical relativistic quantum fields (section 2).  Although the scale curvature  field  (the Weylon) stays  in the background it  may become important for  stabilizing (quantum) fluctuations of the scalar field, if one starts tgo investigate such problems more closely. Here we can, for most of our purposes,  {\em pass to integrable Weyl geometry}.\footnote{In four space-time dimensions  the collection of quadratic curvature terms  then reduces to   $L_{R^2}= - \alpha _1 R^2 - \alpha _2 R^{\lambda \nu } R_{\lambda \nu }$ \cite{Lanczos:1938}. The reduced form is   assumed  in \cite[389]{Nieh:1982}, \cite[260]{Smolin:1979}, \cite[1028]{Drechsler/Tann}.  It also covers the simplified expression of the gravitational Lagrangian in Mannheim's conformal gravity built on   $L_{conf}= C_{\lambda \mu \nu \kappa}C^{\lambda \mu \nu \kappa}$, with $C$ the Weyl tensor \cite{Mannheim:2005}.  } 

At many  occasions  also $L_{R^2}$ may  be neglected;\footnote{P. Mannheim  indicates that this may be acceptable only in the medium gravity regime; he considers the conformal contribution to  extremely weak gravity as crucial \cite{Mannheim:2005}. } 
 then the  Lagrangian of  {\em integrable Weyl-Omote-Dirac} (iWOD) gravity reduces effectively to
\begin{equation} L_{iWOD}= - \epsilon _{sig}\frac{\xi^2}{2} |\phi|^2 R  + \epsilon _{sig} \frac{1}{2} D_{\nu } \phi^{\ast} D^{\nu } \phi -  \frac{\lambda }{4}|\phi|^4  + L_m \, .  \label{L_iWOD}
\end{equation} 
That is very close to the Lagrangian used in recent publications on Jordan-Brans-Dicke theory, e.g.   \cite{Fujii/Maeda}. In Riemann gauge it is nearly identical with the  ``modernized'' JBD Lagrangian of the Jordan frame, the only difference being the $|\phi|^4$-term and the {\em explicit} scale invariance of the Lagrangian. In other gauges (frames) the derivative terms of the rescaling function are ``hidden'' in the Weyl geometric
 terms.\footnote{Cf. \cite{Almeida/Pucheu:2014}. The old version of the JBD parameter corresponds to $\omega =\frac{1}{2}\xi^{-2}$. Contrary to what one might think at first glance, (\ref{L_iWOD}) {\em does not stand in contradiction} to high precision solar system obervations, because  the ``scale breaking'' condition for the scalar field by the quartic potential  prefers scalar field gauge (``Einstein frame'') -- see below. }

\subsubsection*{4.3 The dynamical equations of iWOD}
 \addcontentsline{toc}{subsection}{4.3 The dynamical equations of iWOD}
Variation of the Lagrangian with regard to the Riemannian component of the metric leads to  an  Einstein equation very close to the ``classical'' case; but now  the curvature terms  appear in {\em Weyl geometric} form.\footnote{If one varies the Riemannian part of the metric $g$ and the affine connection $\Gamma$ separately (Palatini approach),  the variation of the connection leads to the {\em compatibility condition} (\ref{compatibility}) of Weyl geometry   \cite{Poulis/Salim:2011,Almeida/Pucheu:2014}. That gives additional (dynamical) support to the Weyl geometric structure. Further indications of its  fundamental role comes frome a completely different side,  a $f(R)$ approach enriched by an EPS-like property \cite{Capoziello_ea:EPSetc}.}
 For $\mathcal{L}_{iWOD}$ without further matter terms the modified {\em Einstein equation}   becomes
\begin{equation}  Ric - \frac{R}{2}g =  \Theta^{(\phi)} =  \Theta^{(I)}+  \Theta^{(II)} \, ,\label{massless Einstein equation} 
\end{equation}
where the  right hand side is basically the energy-momentum  $\Theta^{(\phi)}$ of the  scalar field (multiplied by $(\xi |\phi|)^{-2}$). It   decomposes into a term proportional to the metric, $\Theta^{(I)}$, therefore of the character of vaccum energy or ``dark energy'', and another one which behaves  matter-like (compare the special case studied in section 6.2), $  \Theta^{(II)}$:
\begin{eqnarray}   \Theta^{(I)} &=&   |\phi|^{-2}\left(  -  D^{\lambda} D_{\lambda} |\phi |^2  +  \epsilon_{sig}  \xi^{-2}   \frac{\lambda}{4} |\phi|^4    - \frac{\xi^{-2}}{2 }   D_{\lambda}\phi ^{\ast}D^{\lambda} \phi  \right) \, g  \nonumber \\
   \Theta^{(II)}_{\mu \nu} &=&  |\phi|^{-2}\left( D_{(\mu}D_{\nu)}|\phi|^2 +  \xi^{-2}D_{(\mu}\phi^{\ast}D_{\nu)}\phi     \right)  \label{Theta} 
  \end{eqnarray}
The (``ordinary'') summands with factor $ \xi^{-2}$ are derived from the kinematical 
$\phi$-term of the Lagrangian; the other summands arise from a boundary term while varying the modified Hilbert action. Because of the     variable factor  $|\phi|^2$, the boundary term no longer vanishes  like in the classical case.\footnote{\cite[64ff.]{Tann:Diss},\cite[96ff.]{Blagojevic:Gravitation}, \cite[40ff.]{Fujii/Maeda}. } 
The additional term is  often considered as an ``improvement''  of the energy momentum tensor of the scalar field 
\cite{Callan/Coleman/Jackiw}.\footnote{Callan, Coleman, and Jackiw postulated these terms while studying perturbative scattering theory in a weak gravitational field. They noticed that the ordinary energy momentum tensor  of a scalar field does not lead to finite matrix elements ``even to the lowest order in $\lambda $''. The ``improved'' terms lead to finite matrix terms to all orders in $\lambda $ \cite{Callan/Coleman/Jackiw}. }

All terms of the modified Einstein equation of iWOD gravity (\ref{massless Einstein equation}) are  {\em scale invariant},\footnote{Sometimes the scale transformations are called ``Weyl transformations'' in this context, e.g. in  \cite{Blagojevic:Gravitation}.}
although the geometrical structure is richer than  conformal geometry. Of course there arises the question whether such a geometrical framework may be good for physics, without specifying a preferred scale; i.e., before ``breaking'' of scale symmetry. We shall see in the next section that there is a natural  mechanism for such   `breaking', which is not mandatory (at the classical level) on purely theoretical grounds. 

Constraining the variation to  {\em integrable} Weylian metrics leaves no dynamical  freedom for the scale connection;  thus no dynamical equation arises for  $\varphi$.\footnote{The variation of the Riemannian component of the metric can be restricted to Riemann gauge $(g,0)$. Note the analogy to the variation in JBD gravity of the Riemannian metric with regard to the Jordan frame.} Varying with regard to a real scalar field $\phi$, on the other hand, gives a Klein-Gordon type equation with a ``funny'' mass-like term:
\begin{eqnarray} D_{\nu}D^{\nu}\phi + 2(\xi^2 R +  \epsilon_{sig}\lambda |\phi|^2)\phi + \frac{\delta L_m}{\delta \phi} =0 \label{KG equ}
\end{eqnarray} 

In a way, the scale connection $\varphi$ and the scalar field $\phi$ are closely related. It is possible to scale $\phi$ to a constant, then in general $\varphi \neq 0$; on the other hand one can scale $\varphi=0$, then in general $\phi \neq const$.  The 'kinematical' (descriptive) freedom of $\varphi$ is essentially governed by the dynamics of $\phi$. The {\em scalar field} $\phi$, not the scale connection $\varphi$ encodes the additional {\em dynamical degree of freedom} in the integrable (iWOD) case, far below Planck scale.

\subsubsection*{4.4  Ground state of the scalar field}
\addcontentsline{toc}{subsection}{4.4  Ground state of the scalar field}
There are no reasons to assume that $\phi$ represents an elementary  field. Like all other scalar fields of known physical relevance it may characterize an aggregate state. From our context we may guess that it could represent   an {\em order parameter of a collective quantum state}, perhaps a condensate,   of the Weyl field. Such a conjecture has already been stated in
 \cite[263]{Hehl_ea:Kiel_I},  \cite[1096]{Hehl_ea:Progress1989}, and similarly already in
 \cite{Smolin:1979,Nieh:1982}. 
Here we are not interested in details of the dynamics given by its variational Klein-Gordon equation, but mainly in the ground state which may be indicative for the transition to Einstein gravity.

Transition to integrable Weyl geometry is not yet sufficient to get rid of rescaling freedom. A {\em full breaking of scale symmetry} --- like that of any other gauge group ---  contains {\em two} ingredients: 
\begin{itemize}
\item[(a)] effective vanishing of the curvature (field strength) at a certain scale,
\item[(b)]  physical selection of a specific gauge.\footnote{``Physical'' means  a selection  with observational consequences. Mathematically, the selection of a gauge corresponds to the  choice  of a section (not necessarily flat) in the corresponding principle fibre bundle, at least locally (in the sense of differential geometry). \label{fn choice of section} \\[-2.3em]} 
\end{itemize}
Up to now only step (a) has been taken.    (b)  involves a  ground state of the scalar field with respect to the biquadratic  potential given by its gravitational coupling if the scalar field has the chance to govern the behaviour of physical systems serving as ``clocks'' or as mass units (see section 5). 

For field theoretic investigations signature $sig(g)=(1,3)$ is best suited, so that  $\epsilon_{sig}=+1$. Abbreviating the gravitational  terms we get  $L_{iWOD}= \frac{1}{2} D_{\nu } \phi^{\ast} D^{\nu } \phi  -  V_{grav}(\phi)$ with
\begin{equation} V_{grav}(\phi) = \frac{1}{2} \xi^2 |\phi|^2 R + \frac{\lambda}{4} |\phi|^4 \, . \label{quartic potential}
\end{equation}

In  most important cases, scalar curvature $R$ of cosmological models is  
negative.\footnote{The higly symmetric Robertson-Walker models of Riemannian geometry, with warp (expansion) function $f(\tau)$ and constant sectional curvature $\kappa$ of spatial folia, have scalar curvature $_g \hspace{-0.2em}R= - 6\left( (\frac{f'}{f})^2 + \frac{f''}{f} + \frac{\kappa}{f^2}  \right)$ in signature $(1,3)\sim(+---)$. For $\kappa \geq 0$, or at best moderately negative sectional curvature, and accelerating or ``moderately contracting'' expansion, 
$_g \hspace{-0.2em}R < 0$.   }
Thus the effective gravitational potential of the scalar field is  biquadratic and of  ``Mexican hat'' type  with two minima symmetric to zero,  like in  electroweak theory. Here, however, the coefficient of the quadratic term $\frac{\xi^2}{2} R$ is a point dependent function, but may be scaled to a constant.

The scalar field assumes the gravitational potential minimum for
\begin{equation} |\phi_o| ^2 = - \frac{\xi^2 R}{ \lambda } \qquad \mbox{(in reciprocal length units),}  \label{potential condition}
\end{equation}
and the ``funny'' mass  term of the Klein-Gordan equation (\ref{KG equ}) vanishes in the undisturbed ground state. For the moment we have to leave it open, which kind of disturbances might shift  the scalar field away from its potential minimum of (\ref{potential condition}).  

Of course, there is a scale gauge in which $|\phi_o|$ assumes constant values. We call it the {\em scalar field gauge} (of Weyl geometric gravity). 
 Starting from any gauge $(g,\varphi)$ of the Weylian metric, just rescale by $\Omega:= C^{-1}|\phi_o| $ with any constant $C$. Because of it having scale weight $-1$, the norm of the scalar field then becomes $|{\phi}_o(x)|\doteq C$ in inverse length units;  equivalently in energy units 
\begin{equation} |{\phi}_o(x)| [\hbar c]\doteq  C \hbar c =:|\phi_{c}| \;  \label{scalar field gauge}
\end{equation}
  with some constant energy value $|\phi_{c}|$. The  {\em dotted equality} $\doteq$ expresses that the relation is no longer scale invariant but holds in a specific gauge only, here in the scalar field gauge.

With $C$ such that $\xi^2 C^2 = (8 \pi G)^{-1}[\frac{c^4}{\hbar c} ]$ ($G$ gravitational constant)  the coefficient of the iWOD-Hilbert term (\ref{L_iWOD}) goes over into the one of Einstein gravity. Then
\begin{equation} |\phi_{c}| = \xi^{-1} \left( \frac{\hbar c^5 }{8 \pi G}  \right)^{\frac{1}{2} }= c^2 M_{pl} = E_{pl}\quad \mbox{(reduced Planck energy)}  , \label{compatibility Einstein gr}
\end{equation}  
and the coupling constant  $\xi^2$ turns out to be basically a squared hierarchy factor  between the scalar field ground state in energy units and Planck energy $E_{pl}$.

\subsubsection*{4.5 Scale invariant observables and a new look at `dark energy'}
\addcontentsline{toc}{subsection}{4.6 Scale invariant observables and a new look at `dark energy'}
It is easy to extract a {\em scale invariant observable magnitude}  $\hat{X}$ from a scale covariant field $X$ of weight $w(X)=k$. One only has to form the  proportion with regard to the appropriate power of the scalar field's norm 
\begin{equation} \hat{X}:= X / \, |\phi|^{-k}  = X |\phi|^k\, ;  \end{equation}
then clearly $w(\hat{X})= 0$.

 Scale invariant magnitudes $\hat{X}$  are directly indicated, up to a globally constant factor in scalar field gauge, i.e., the gauge in which  $|\phi_o| \doteq const$.\footnote{In  \cite{Utiyama:1975I}   $\phi$ is therefore called a ``measuring field'';  cf. \cite{Scholz:AdP}.} 
Conceptually the problem of scale invariant magnitudes is solvable, even with  full  scaling freedom, but there are physical effects  which lead to  actually breaking scale symmetry. Atomic ``clocks'' and ``rods'' (atomic distances) express a preferred metrical scale. They  stand in good agreement with other periodic motions of physics on different levels of magnitude.

The ordinary energy-momentum terms with scale covariant derivatives of $\phi$ in (\ref{Theta}) get  suppressed by the inverse squared hierarchy factor $\xi^{-2} < 10^{-32}$ (see section 5.3). Only the $\lambda $-term  corresponding to the old cosmological term survives because   it is  of fourth order   in $|\phi |$ and $|\phi |$ is sufficiently large.   In the ground state  $|\phi|^2$  can be expressed  in terms of the scalar  curvature,
 (\ref{potential condition}). Then the energy-momentum of the scalar field  simplifies to (remember: $g=(g_{\mu \nu})$ stands for  the whole metric): 
\begin{eqnarray}   \Theta^{(I)} &\approx &   \left(  -  \frac{R  }{4}  -   |\phi|^{-2} D^{\lambda} D_{\lambda} |\phi |^2 \right)  g  =: \Lambda  \,g   \label{vacuum energy} \\
 &\approx&   \left(  -  \frac{R  }{4}  -   R^{-1} D^{\lambda} D_{\lambda} R \right)  g      \nonumber   \\
   \Theta^{(II)}_{\mu \nu} &\approx &  |\phi|^{-2} D_{(\mu}D_{\nu)}|\phi|^2 =  R^{-1} D_{\mu} D_{\nu} R  \label{"dark matter"}
\end{eqnarray} 
This expresses a peculiar back-reaction of curvature (gravity) on itself via the scalar field, which is not present in  Einstein gravity. Of course it als complicates the dynamical eqauations.\footnote{In general, the order of the Einstein equation is raised to four, although in Weyl gauge it remains of second order!}

Taking traces on both sides of the (iWOD) Einstein equation shows that  in
 the {\em matter free} case, $L_m=0$,
\begin{equation} |\phi|^{-2} D^{\lambda} D_{\lambda} |\phi |^2 \approx 0 \, ,  \label{vacuum Theta I}
\end{equation}  
and the vacuum Einstein equation can be written in a trace free form: 
\begin{equation} Ric - \frac{R}{4}g \approx \Theta^{(II)} \label{trace free Einstein equ} 
\end{equation} 

These identities  signal a  remarkable change in comparison with  Einstein gravity and its  problems with the cosmological constant. $\Theta^{(I)}$  represents a functional equivalent to the traditional  ``vacuum energy'' term, but here it is due to the scalar field. The coefficient $\Lambda $ in (\ref{vacuum energy}) {\em depends  on the geometry of iWOD gravity and thus, indirectly, on the matter distribution}. 
Moreover,   $\Theta^{(II)}$  is an additional contribution to the energy momentum of the scalar field (\ref{"dark matter"}). Perhaps we can expect that some of the effects ascribed  to  {\em dark matter} may be due to  it.

\subsubsection*{4.6 A first try of connecting to electroweak theory}
 \addcontentsline{toc}{subsection}{4.5 A first try of connecting to electroweak theory}
It seems tempting to consider the electroweak energy scale  $v$ as a candidate for the value of the gravitational scalar field in scalar field gauge, 
\[ |\phi_{c}| = v \approx 246 \, GeV \, . 
\] 
 In this case,  the value of the hierarchy factor would be  $\xi =   \frac{E_{pl}}{ v} \sim 10^{16} $.

With
 \begin{equation} \lambda \sim 10^{-56}, \label{lambda}
 \end{equation} 
 the value of the scalar field's ground state is located, by (\ref{potential condition}),  at the electroweak scale:\footnote{Here $ |R| \dot{\sim} H^2$ with $H=H_1 \approx 7.6 \cdot 10^{-29}\, cm^{-1}$, respectively $\hbar c\, H \approx  1.5 \cdot 10^{-33} \, eV \sim 10^{-32}\, eV$.  In  section 6 we find good reasons to consider $R \doteq 24 H^2$ (\ref{stable Einstein universe}). 
 \label{fn H in ev}}
 \begin{equation} [\hbar c] |\phi_o| = \hbar c  \frac{\xi \sqrt{|R|}}{\sqrt{2  \lambda}} \;  \dot{\sim} \; 10^{16-33+28}\, eV \sim 10^{11}\, eV\, , \quad  |\phi_o| \, \dot{\sim} \, 10^{16}\, cm^{-1} \label{value of phi}
 \end{equation}
 
 This observation  indicates a logically possible connection between Weyl gravity (iWOD) and electroweak theory, although the order of magnitude  of $\lambda$ looks quite suspicious. 
 In the next section  we explore   a  related, but more convincing  { transition}   which gives up the idea that the gravitational scalar field might be immediately identified with the Higgs field.  
Our goal is  to find out whether there is a chance for the  scalar field  to  determine the rate of clock ticking and  to  influence the units of mass by some relation to the electroweak theory.

\subsection*{5. A bridge between Weyl geometric gravity and ew theory}
  \addcontentsline{toc}{section}{5. A bridge between Weyl geometric gravity and ew theory}
Let us try  to explore whether   the Weyl geometric setting may  contribute  to conceptualizing the ``generation of mass'' problem of elementary particle physics. Mass is the charge of matter fields with regard to the
 inertio-gravitational field, the affine connection of spacetime. In flat space, and thus in special relativity, that  may   fall  into oblivion  because there the affine connection is hidden under  the pragmatic form of  partial derivatives. The exercise of importing standard model fields to ``curved spaces'', i.e., Lorentzian or 
Weyl-Lorentzian manifolds,   is  conceptually helpful  even if it is done   on a classical level as a first step. Using  Weyl geometry  seems all the more appropriate, as nearly all of the Lagrange terms of the standard model of elementary particle physics (SM) are  already conformally invariant. The only exception is the quadratic term fo the Higgs field, $\frac{\mu^2}{2}|\Phi|^2$, with the dimensional factor $\mu^2$. By means of the gravitational scalar field it can easily brought into a scale covariant form of the correct weight.\footnote{\cite{Nishino/Rajpoot:2004,Nishino/Rajpoot:2009,%
Meissner/Nicolai,Bars/Steinhardt/Turok:2014}}

\subsubsection*{5.1 Importing standard model fields to IWG}
\addcontentsline{toc}{subsection}{5.1 Importing standard model fields to IWG}
Most contributions to the special relativistic  { Lagrange density $L_{SM}(\psi)dx$ of the standard model} of elementary particles (SM) are invariant under dilations in Minkowski space. 
Dilational invariance is closely related to unit rescaling, but not identical.
Assigning Weyl weight $w=-d$ to a a field $\psi$ of dilational weight $d$ (ofte called ``dimension'') gives an invariant Lagrangian density under global unit rescaling in special relativity.\footnote{Under the active dilation of Minkowski space $x\mapsto \tilde{x}=\Omega x $ ($\Omega>0$ constant) a field $\psi$ of dilational weight $d$ transforms by $\psi(x) \mapsto \Omega^d \psi(\Omega^{-1}x)$ \cite[682ff.]{Peskin/Schroeder}. Invariance of the action $S=\int L(\psi)dx$ holds if $\int L(\psi(x))dx= \int \tilde{L}(x)\Omega^{-4}dx$. That is the case if and only if $\tilde{L}=\Omega^4 L$, thus $d(L)=4$ and $w(L)=-4$ for Lagrangians  invariant under dilations. Rescaling  $\eta = diag(1,-1,-1,-1)$ by  $\eta \mapsto \tilde{\eta}= \Omega^2 \eta$ leads to $L \sqrt{|det \, \eta|}= \tilde{L} \sqrt{|det\,\tilde{\eta|}}$ and thus to a scale invariant Lagrange density. \label{fn dilations}}
Unit rescaling can be made point dependent, if the fields can be generalized to the Weyl geometric framework.

 An energy/mass scale is introduced into the SM by the Lagrangian of the Higgs-e.a. mechanism.\footnote{Spelt out, Brout-Englert-Guralnik-Hagen-Higgs-Kibble ``mechanism''.}
 One usually assumes that the Higgs field is an {\em elementary} scalar field with values in an
 isospin-hypercharge representation $(I,Y)=(\frac{1}{2},1)$   of the electroweak group $G_{ew}= SU(2)\times U(1)$.\footnote{With the ordinary Gellmann-Nishijima relation $Q=I_3 + \frac{1}{2}Y$ usually assumed in the literature.  Drechsler uses a convention for $Y$, such that $Q=I_3 + Y$.} At least two generations of particle physicists have been working in the expectation that this scalar field is carried by a massive boson of rest mass at the electroweak level ($\sim 100\, GeV$). Experimenters at the LHC have finally found striking evidence for such a boson with mass $m_H \approx 125 - 126 \, GeV$ \cite{ATLAS:Higgs2012,CMS:Higgs2012}.

Without going too much into detail, it can be stated   that all the fields and differential operators of the standard model Lagrangian can be imported into  Weyl 
geometry.
The most subtle question is the representation of the Weylian covariant derivative for fermionic fields.\footnote{Here we are mainly concerned with the Higgs sector, so we do not need to consider all details of the Weyl geometric version of $\mathcal{L}_{SM}$. For a  complete formulation  see \cite{Nishino/Rajpoot:2004,Nishino/Rajpoot:2009}, similarly, from a purely conformal view  \cite{Meissner/Nicolai};  for the ew sector  see \cite{Drechsler:1991,Drechsler/Tann,Scholz:AdP}. The scalar field and scale connection (Weylon) sector is introduced in \cite{HungCheng:1988}. A short discussion of the local bundle construction  in Weyl geometry is given  by \cite{Drechsler/Hartley:1994}; for the Riemannian case see, e.g.,  \cite[chap. 19]{Frankel:Geometry}.} 

The kinetic term of the special relativistic Dirac action 
$ \frac{i}{2}(\psi^{\ast}\gamma^o \gamma^{\mu}\partial_{\mu} \psi- (\gamma^{\mu}\partial_{\mu} \psi)^{\ast} \gamma^o \psi) $
is  conformally invariant if $\psi$ is given the scaling weight $w(\psi)=-\frac{3}{2}$. After orthogonalizing the Levi-Civita connection by introducing tetrad coordinates (in the tangent bundle)
it is  locally given  by a 1-form $\omega$ with values in $so(1,3)$. Using the appropriate spin representation 
 it can be ``lifted'' to spinor fields.\footnote{In 1929, Weyl and Fock noticed  independently that in this construction a point dependent phase can be chosen freely without affecting observable quantities. That implied an additional $U(1)$ gauge freedom and gave the possibility to implement  a $U(1)$-connection \cite{Scholz:FockWeyl}. Their original proposal to identify the latter with the electromagnetic potential was not accepted because all fermions would  seem  to couple non-trivially to the electromagnetic field. \cite{Pawlowski:1999} gives the interesting argument that in electroweak theory  the hypercharge field can be read as operating on the spinor phase, exactly like Weyl and Fock had proposed for the electromagnetic field \cite{Weyl:1929,Fock:1929}. }
  In this way  the Dirac action  on ``curved'' Lorentzian spaces acquires the form 
\begin{equation} \frac{i}{2}(\psi^{\ast}\gamma^o\gamma^{j}\nabla_j \psi - ( \gamma^{j}\nabla_j\psi)^{\ast} \gamma^o\psi)  \; ,\label{Dirac Lorentz}
\end{equation}
where the latin indices $i, j, k \ldots$ indicate tetrad coordinates, $\gamma^j$ constant, standard Dirac matrices and  $\nabla_j$ (here) the covariant spinor derivative. Notation here: $\psi ^{\ast}= ^t  \hspace{-0.3em} \overline{\psi} $,  \quad $\overline{\psi}$ complex conjugate, $^t$ transposition.
All this can be done globally if  the underlying spacetime manifold $M$ is assumed to be {\em spin}, otherwise only 
locally.\footnote{$M$ is {\em spin}, iff it admits a global $SL(2,\mathbb C )$ bundle; then the  Dirac operator can be defined globally, otherwise only locally (in the sense of differential geometry). A sufficient criterion is $H_2(M, \mathbb Z_2)=0$.}
The action is conformal invariant and is used in conformal approaches to   gravity and SM fields  \cite[85]{Birrel/Davies:QFT}.\footnote{Thanks to P. Mannheim for insisting on this point; cf.  \cite[fns. 20, 21]{Mannheim:2005}. It is important for clarifying  the specific difference between the conformal Dirac action and its Weyl geometric twin.}

For different choices of the  representative of the metric, the conformal approach refers to  {\em different} affine connections, but uses   scale invariant  Lagrangians and equations. In the Weyl geometric approach, on the other hand, rescaling does {\em not change} the affine connection and covariant derivative (see sect. 2.1, eq. (\ref{Levi-Civita})). Therefore  the   
`orthogonalized'  Weyl geometric  connection  $\omega =(\omega^i_{\,j})$, written as 1-form with values in $so(1,3)$,   contains a contribution of the scale connection $\varphi$, $\omega= _g{\hspace{-0.35em}}\omega + _{\varphi}{\hspace{-0.2em}}\omega$ ($_g{\hspace{-0.05em}}\omega$ the orthogonalized Levi-Civita connection of $g$).\footnote{$_{\varphi}{\hspace{-0.05em}}\omega$  is  of the form $(\varphi_i  \eta_{jk} - \varphi_j \eta_{ik}) \vartheta^k = \omega_{ijk}\vartheta^k$, where  $\{ \vartheta^i\}$ denotes the selected  coframe basis,  $\eta$  the Minkowski metric, and latin indices $i, j$ of $\varphi$ indicate its coframe coordinates. \cite[eq. (2.16)]{Drechsler/Hartley:1994}, \cite[eq. (4.39b)]{Blagojevic:Gravitation}). \label{fn phi_omega}} 
This contribution is a {\em specific attribute} of the Weyl geometric  coupling of the scale connection to  spinor fields,  while the usual gauge interaction vanishes (see below). $_{\varphi}{\hspace{-0.01em}}\omega $ takes care for the spin connection being {\em unchanged} under rescaling. Without it  the Audretsch/G\"ahler/Straumann consideration on streamlines of the WKB approximation could not hold indepedendently of the scale gauge (section 2.3).

Finally, the scale covariant derivative for Dirac spinors becomes \begin{eqnarray} D_{\mu} \psi &=& \partial_{\mu} \psi + \frac{1}{4}[\gamma^i, \gamma^j ]\, \omega_{ij{\mu}}\, \psi - \frac{3}{2} \varphi_{\mu}  \psi \nonumber  \\ 
   /\hspace{-0.7em}D \psi &=&  [ \hbar c] \, \gamma ^{\mu}  D_{\mu} \psi    \, , \label{Dirac operator}
   \end{eqnarray} 
   with $w(\gamma^{\mu}) =-1$ and  $w(/\hspace{-0.7em}D \psi ) = -\frac{5}{2} $.\footnote{$\{\gamma^{\mu}, \gamma^{\nu} \}= 2g^{\mu \nu}$ implies  $w(\gamma^{\mu})=-1$, while by the same reason $w(\gamma_j)=w(\gamma^j)=0$.}
      The kinetic term of the action is  formed analogous to (\ref{Dirac Lorentz}). In simplified form (passing over the chiral decomposition of the spinor fields) the massless Dirac action and the corresponding Yukawa mass term can be written as:
   \begin{eqnarray}  
   L_{\psi}   &=&  \frac{i}{2} (\psi^{\ast} {\gamma}^o \, /\hspace{-0.7em}D \psi - (/\hspace{-0.7em}D \psi)^{\ast} {\gamma}^o \psi) \label{Dirac action Weyl geo} \\
     L_{Y}   &=&  - \mu_{\psi} |\phi| \, \psi^{\ast}{\gamma}^o\,  \psi \nonumber
     \end{eqnarray}
     $L_{\psi}$ and $L_Y$ are  of weight $-4$. Thus in the Weyl geometric theory not only the massless Dirac field but  also the {\em massive} one has a {\em scale invariant} Lagrangian density.  Due to  hermitian symmetrization  the real valued gauge couplings  $- \frac{3}{2} \varphi_k  \psi $ from (\ref{Dirac operator})  cancel in (\ref{Dirac action Weyl geo}).\footnote{Even for the non-integrable case this cancelling takes place \cite[81, ex.1]{Blagojevic:Gravitation}, \cite{Mannheim:2014}, already noted by \cite[440]{Hayashi_ea:1977}.  
   Although there is  no  gauge coupling of the Yang-Mills type,  the scale covariance term $- \frac{3}{2} \varphi_k  \psi $  has to be retained for  consistency reasons (in the Lagrangian and the resulting Dirac equation). Dynamical effects  of the scale connection result only from $_{\varphi}{\hspace{-0.1em}}\omega$ ($\rightarrow$ fn. \ref{fn phi_omega}).}

We   rebuild crucial aspects of the Higgs field in our framework by  extending the scalar field of iWOD gravity to  an electroweak bundle of appropriate maximal weight  for $G_{ew}$, $(I,Y)=(\frac{1}{2},1)$. The scalar field turns into  a field $\Phi $  with  values in a point dependent representation space isomorphic to  $\mathbb C^2$,%\footnote{Mathematically speaking, $\Phi $ is a section in an  associated vector bundle with $(I,Y)=(\frac{1}{2}, 1)$ of the electroweak principal bundle.}
\begin{equation} \Phi (x) = (\phi_1(x), \phi_2(x)) \, . \label{Phi}
\end{equation}

\subsubsection*{5.2 Two steps in the geometry of symmetry breaking}
  \addcontentsline{toc}{subsection}{5.2 Two steps in the geometry of symmetry breaking}
The usual ``mechanism'' for electroweak symmetry breaking on the classical level consists of two components. 
\begin{itemize}
\item[(I)] By a proper choice of  $SU(2)$ gauge $ \Phi (x)$ is  transformed into a  ``down'' state at every  point; $\Phi (x) = (0, h(x))$, with complex valued $h(x)$.
\item[(II)] In the ground state of $\Phi$, its (squared) norm, physically spoken the expectation value  $<\Phi ^{\ast}\Phi>$,  is assumed to lie in a minimum of  a  quartic (``Mexican hat'') potential. We write   $\Phi_o = (0, h_o(x))$. In the classical Higgs theory its norm is a constant, $|h_o(x)|= const = v$.
\end{itemize}

In the  physics literature (I) is considered as a {\em spontaneous breaking} of the $SU(2)$ symmetry. This happens without reducing the symmetry of the Lagrangian.  For step (II) in the usual understanding of the Higgs procedure, a mass scale is introduced into the otherwise  (globally) scale invariant Lagrangian of the  standard model ; i.e. scale symmetry is  {\em explicitly broken}. In our context, we have to reconsider the last point. But before we do so, we shall have a short look at the features of the spontaneous breaking in step (I). This will help us in transforming step (II)  into a breaking of the spontaneous type, which we want to adress in section 5.3. 

 The first step presupposes the ability to specify  ``up'' and ``down'' states with regard to which the ``diagonal'' subgroup of $SU(2)$ with generator $\sigma_3=\frac{i}{2}\,diag(1,-1) $ is defined. Otherwise the $U(1)$ subgroup could be any of infinitely many conjugate ones.\footnote{There are infinitely many {\em maximal tori} subgroups, all of them can serve with equal right as ``diagonal'' (Cartan) subgroup. The ``localization'' (in the sense of physics) allows to make the selection point dependent.} Stated in more physical terms: How do we know in which ``direction'' (inside $\mathbb C^2$) the 3-component of isospin has to be considered? This question, already important in special relativistic field theory, becomes pressing in a consequently ``localized'' (in the physical sense) version of the theory; i.e., in passing to general relativity. 
 
 In the following we shall consider the Weyl geometrically extended Higgs field $\Phi$ and investigate  whether the (complex valued) down state component $h(x)$ of the Higgs field may be related to the gravitational scalar field $\phi(x)$.

It seems natural to assume that the {\em ground state of the electroweak vacuum field $\Phi(x)$ defines the down state of the vacuum representation} of the electroweak group, $(I,Y)=(\frac{1}{2},1)$, at every point $x$. Thus a subgroup  $U(1)_o \subset SU(2)$ is specified as the isotropy group (fix group) of the complex ray generated by $\Phi(x)$ at each point. It  singles out the  $I_3$ and charge eigenstates in all  associated representations of $G_{ew}$, and thus for the elementary fields.  In consequence, an {\em adapted basis in each of the representation spaces} can be chosen at every  point, such that wave functions of the up/down states get their usual form. The scalar field, e.g.,  goes over into the form of the preferred electroweak gauge (often called ``unitary gauge'')
\begin{equation} \Phi(x)= (0, h(x))\, , \label{unitary gauge}
\end{equation}
and the only degrees of freedom  for $\Phi$ are those of  $h$, a complex valued field. 

In this way the Higgs field specifies,   at each point $x\in M$,
a subgroup $U(1)_o \subset SU(2)$, mathematically a maximal torus of $SU(2)$, in  $G_{ew}= SU(2)\times U(1)$. The eigenspaces of $U(1)_o$  are the $I_3$ eigenstates of the corresponding isospin representation spaces with $I\in \{\frac{1}{2}k | \, k\in \mathbb N   \}$.
In physical terms, the ew dynamics  is ``informed''  by the Higgs field  how the weak and the hypercharge group (or Liealgebra) are coordinated in the generation of electric charge, also for other (fermionic) representation spaces.\footnote{Experiment has shown  that for left handed elementary fields (and for  the ``vacuum'')   
 $I=\frac{1}{2}  $.   At any point of spacetime the charge eigenstates of left handed elementary matter fields are specified by the dynamical structure of the vacuum as the eigenstates ($I_3= \pm \frac{1}{2} $) of $U(1)_o $ and  $Q=I_3+\frac{1}{2}Y$. $(I,Y)= (\frac{1}{2}, -1  )$ for (left-handed) leptons,   $(I,Y)= (\frac{1}{2}, \frac{1}{3}  )$ for (left-handed)  quarks, and  $(I,Y)=(\frac{1}{2},1)$  for the ``vacuum''. For right handed elementary fields the isospin representation is trivial, $(I,Y)=(0,2Q)$.} 
 In this sense, the electroweak symmetry does not treat every maximal torus ($U(1)$) subgroup of the  $SU(2)\subset G_{ew}$ equivalent to any other. The Higgs field, encoding an important part of the physical vacuum structure, seems to be crucial for the distinction.

In this way the Higgs-e.a. mechanism,  can be imported to the general relativistic framework.  The whole structure can still be transformed under  point dependent $SU(2)$ operations without being spoiled, i.e., it may be gauge transformed.\footnote{``Active'' gauge transformations operate on the whole setting of $\Phi(x), U(1)_o$ and the corresponding frame of up/down bases --- similar to the diffeomorphisms of general relativity, considered as gauge transformations; they carry   the metrical structure with  them. The active transformations can be countered by ``passive'' ones which, in mathematical terminology, are nothing but an adapted change of the trivialization of a principle fibre bundle and accompanying choices of standard bases ($I_3$ eigenvectors) in the associated representation spaces. After a joint pair of active and passive gauge transformations the wave functions expressed in ``coordinates''  remain the same. \label{fn active passive}}
 And even more importantly, if a   $\mathfrak{su}(2)$ or $\mathfrak {g}_{ew}$ connection
 of nonvanishing curvature, i.e., an electroweak field,  is present,\footnote{Curly small letters like $\mathfrak{su}(2)$ and $\mathfrak {g}_{ew}$  denote the Liealgebra of the corresponding groups.} 
 it is not reduced to one of vanishing curvature by the pure presence of the scalar (Higgs) field. In that respect, {\em  gauge symmetry remains intact}  in the sense  of both automorphism structure and dynamics.

 The metaphor of ``breaking''  gauge symmetries has been  discussed broadly, often critically,  in   philosophy of science, cf. \cite{Friederich:gaugesymmetry_breaking,Friederich:2014}. 
It did not pass without objection among physicists either, e.g., \cite{Drechsler:Higgs}. For an enlightening historical survey of the rise of the electoweak symmetry breaking narrative and its important {\em heuristic and systematic role} see \cite{Borrelli:Rosetta,Karaca:Higgs}.  
From our point of view, it  does  not seem a particularly happy choice to speak of  ``breaking'' the $SU(2)$ symmetry at this
 stage.
But it is true that the physical specification of the $U(1)_o$ subgroup (maximal torus) in $SU(2$) by the scalar field allows to introduce standard sections ($I_3$ bases) and preferred  trivializations  of the representation bundles, corresponding to step (b) in the characterization of section 4 (above footnote \ref{fn choice of section}). In this sense, the otherwise free choice of a trivialization is ``broken'', and  one can say that a full reduction of the electroweak symmetry, which presupposes vanishing of the curvature (the  field strength)  is {\em foreshadowed} by the presence  of the scalar field. In such a sense there is no problem with the language of ``spontaneous breaking'' of symmetries.

A {\em full breaking} of the dynamical symmetry will be accomplished when, in addition to a preferred gauge choice (trivialization), the physical conditions for an effective vanishing of the $SU(2)$  curvature component are given (step (a) in  section 4.4). That is the {\em result of the gauge bosons acquiring mass},  rather than the origin and explanation of mass generation, although the mass splitting of the fermions is ``foreshadowed'' by the physical choice of $U(1)_o$ subgroup (the ``$I_3$ direction'' in more physical terms). This agrees well with S. Friederich's convincing discussion, including quantum field aspects,  of the Higgs mechanism in \cite{Friederich:2014}.
We come back to this point in a moment. 

The second aspect  of the usual ew symmetry breaking scenario,    (II) in the characterization above,   consists of  reducing the underdetermination of the (squared) norm of $\Phi$, respectively the vacuum expectation value of $\Phi^{\ast}\Phi = |h|^2$.
In the ordinary Higgs-e.a. mechanism that is achieved by ad hoc  postulating a quartic potential of ``Mexican hat'' type for the Higgs field. In the  iWOD approach, a similar potential for the gravitational $\phi$ is {\em naturally} given by  (\ref{quartic potential}), with  ground state in (\ref{potential condition}). It remains to be seen whether the Higgs potential can be related to it in a mathematically and physically convincing way.

Crucial for the Higgs-e.a. mechanism is the fact that covariant derivative terms of the scalar field in ew theory (the ew bundle) lead to {\em mass-carrying Lagrange  terms} for the gauge fields, which are nevertheless {\em consistent with the full gauge symmetry}. This is, of course, just so in the ew-extended iWOD model.  
The kinematical term of the scalar field becomes now
\begin{eqnarray}  L_{\Phi} &=&  \frac{1}{2}\tilde{D}_{\nu}\Phi^{\ast} \tilde{D}^{\nu}\Phi \, , \qquad  \label{L_Phi}\\
 \tilde{D}_{\mu}\Phi  &:=& (\partial_{\mu }-\varphi_{\mu} + \frac{1}{2}{g} W_{\mu} +  \frac{1}{2}{g}' B_{\mu}) \Phi  \, ,   \nonumber %\label{ew dynamical derivative Phi}
\end{eqnarray} 
where the $ W_{\mu}$ and $ B_{\mu}$ denote the connections in the $\mathfrak{su}(2)$ and $\mathfrak{u}(1)$ component of the electroweak group respectively. 
The ew covariant derivative terms of (\ref{L_Phi}) lead to scale covariant formal mass terms for the ew bosons.\footnote{$ \frac{1}{4} {g}^2 |\Phi|^2 W_{\mu}W^{\mu} \qquad \mbox{and} \qquad   \frac{1}{4}  {g}'^2 |\Phi|^2 B_{\mu}B^{\mu} $.}  
After the settling of $\Phi$ in a ground state, $\Phi_o =(0, h)$, we hope to find an appropriate scale in which $h\doteq const = v$ (``{\em Higgs gauge}''). Then,   after a  change of basis (Glashow-Weinberg rotation), the formal mass terms turn into explicit ones
\begin{equation} m_W^2 = \frac{g^2}{4}v^2\, , \qquad m_Z^2 = \frac{g^2}{4 \cos{\Theta }^2}v^2  \; ,  \label{ew boson masses} 
 \end{equation}
with  $\cos \theta = {g} \,(g^2 + g'^2)^{-\frac{1}{2} } $  like in special relativistic field 
theory.

 Already in the special relativistic case it is much more difficult to establish $G_{ew}$ invariant and scale invariant  Lagrangian densities for the fermionic fields, in particular with regard to the  mass terms.\footnote{Decomposition in chiral (left and right) states and the transformation on mass eigenstates for quarks (Cabibbo-Kobayashi-Maskawa (CKM)  matrix)  and leptons (Maki-Nakagawa-Sakate (MNS) matrix) have  to be taken into account.  The Yukawa Lagrangian for the fermions are  simplest, if written in unitary gauge 
(\ref{unitary gauge}), but are   gauge invariant, cf. fn
 \ref{fn active passive}. } 
The transfer to the Weyl geometric context is  a smaller problem, once that has been 
achieved.\footnote{\cite{Drechsler:Higgs,Nishino/Rajpoot:2009,Scholz:AdP}, cf.  \cite{Meissner/Nicolai}. }
Basically one has to adapt the Dirac operator  (\ref{Dirac operator}) to the Weyl geometrical context. 
In simplified form, the resulting Lagrangian for  electrons  can be written as
 \begin{equation} L_e = \frac{i}{2} (\psi_e^{\ast} {\gamma}^o \, /\hspace{-0.7em}D \psi_e - (/\hspace{-0.7em}D \psi_e)^{\ast} {\gamma}^o \psi)   - \mu_{e} |\phi| \, \psi_e^{\ast}{\gamma}^o \psi_e  \, ,  \label{fermionic Lagrangian }
 \end{equation} 
 with  $\mu_e$ the coupling coefficient for the interaction of $\phi$ and the electron  field. 
 
The fermions and the weak gauge bosons acquire their mass  from their interactions with Higgs field in its ground state $\Phi_o$.   For the electron
\begin{equation} m_e = \mu_e [h c] |\Phi_{o}| \doteq \mu_e v \, .  \label{electron mass}
\end{equation}
Once the weak bosons have acquired mass $m_w$, the  range of the exchange forces mediated by them is limited to the order of $l_w = \frac{\hbar c}{m c^2} \sim 10^{-16}\, cm$. At distances  $d \gg l_w$ the curvature of the weak component in the group $G_{ew}= SU(2)\times U(1)$ vanishes effectively,  the weak gauge connection can be ``integrated away'', and the symmetry can be effectively reduced to $U(1)$.  As a result, {\em electroweak symmetry is broken down to the electromagnetic subgroup}. That happens  {\em because of the mass acquirement of the weak bosons}  -- {\em not the other way round}. In this respect  the physical interpretation of our stepwise  reduction  deviates slightly from the standard account, although the basic structure of the Higgs-e.a. mechanism has been taken over  in most respects. 

We still have to  face the fact that in the Weyl geometric setting even the ground state $\Phi_o$ has to be  scale covariant of weight $w(\Phi)=-1$, just like the gravitational scalar field $\phi$. We therefore have to look for a modification of the classical Higgs potential, adapting it to Weyl geometry and forging a bridge, a ``transition'', between the electroweak (Higgs) scalar field and the gravitational one. 

\subsubsection*{5.3 Intertwinement of  the Higgs field with gravity's scalar field}
\addcontentsline{toc}{subsection}{5.3 Intertwinement of  the Higgs field with gravity's scalar field}
The usual Higgs ``mechanism'' works with a Lagrangian of the form
\begin{equation}
\mathcal{L}_{\Phi} = \quad \;  {(\frac{\mu^2 }{2}|\Phi|^2 - \frac{\lambda}{4} |\Phi|^4 + \frac{1}{2} D_{\nu}\Phi^{\ast}D^{\nu}\Phi + \ldots) \sqrt{|det\,\eta|} } \,,\label{L_h}
\end{equation}
with $\eta$ the Minkowski metric,  $|\Phi|^2 = \Phi^{\ast}\Phi$, and $\mu^2 , \lambda$ the effective values for the quadratic and quartic coefficients of the SM Lagrangian at the ew energy level.
The coefficient of the quadratic term  $\frac{\mu^2 }{2}$ is dimensionful and of type energy/mass squared. In our conventioin it would correspond to a quantity of scale weight $w(\mu^2)=-2$. Formally this Lagrangian bares a close resemblance to the one of the Weyl geometric  gravitational scalar field 
\begin{equation}
\mathcal{L}_{\phi} = {( - \frac{1}{2}\xi^2 |\phi|^2 R -  \frac{\lambda}{4} |\phi|^4 + \frac{1}{2} D_{\nu}\phi^{\ast}D^{\nu}\phi + \ldots) \sqrt{|det\, {g}|} } \, . \label{L_V_phi}
\end{equation}
But we have seen  that a direct identification is impossible because of the empirical constraints for the coupling coefficients.\footnote{Moreover, a closer look at galactic and cluster dynamics may speak in favour of introducing another form of the gradient term $L_{\phi}$, e.g. the cubic one of a weylianized AQUAL theory. } 

In order to make  (\ref{L_h}) locally scale invariant,
we first  replace the definite mass value $\mu$ by a scale covariant quantity  which, for the sake of local scale covariance, has to be a scale covariant scalar function  with real or complex values. Nevertheless a  preferred scale indicating the level of  electroweak energy $v = 246\, GeV$ for the expectation value of $\Phi$,\footnote{More precisely we deal here with the square root of the expectation value  $<\Phi^{\ast},\Phi>$  abbreviated by $|\Phi|^2$.}
 clearly a constant relative to the definitions of measurement units, has to arise naturally. For that we need some kind of ``spontaneous breaking'' of the scale symmetry.

In section 4.4 we have observed that the gravitational scalar field $\phi$ shares features of such a spontaneous breaking  by its coupling to the Weyl geometric scalar curvature (\ref{potential condition}),  analogous to criterion (I) in section 5.2. In our context it   seems very natural to consider the hypothesis that the  Higgs field acquires it preferred (``broken'') expectation value, and thus its mass, by its coupling to the gravitational scalar field. 

The simplest form to achieve this  is to assume  a biquadratic potential
\begin{equation}
V_{bi}(\phi, \Phi) = \frac{\lambda}{4}(|\Phi|^2 - \alpha^2 \phi^2)^2 + \frac{\lambda'}{4} \phi^4 \, , \label{V_bi}
\end{equation} 
in addition to the modified 
 Hilbert term from (\ref{L_V_phi}). Simplifying   $|\Phi|^2=h^2$  according to (\ref{unitary gauge}),  the full gravitational  potential becomes 
 \begin{equation}
 V(\phi, h)= \frac{1}{2}\xi^2 |\phi|^2 R +  \frac{\lambda}{4}(h^2 - \alpha^2 \phi^2)^2 + \frac{\lambda'}{4} \phi^4 \, \label{V(phi,h)} \,.
 \end{equation}
And the  scalar field part of the Lagrangian $\mathcal{L}_{\phi \Phi} =  L_{\phi \Phi} \sqrt{|det\, {g}|}$ is 
 \begin{equation}
 L_{\phi \Phi} = - \frac{\xi^2 }{2}|\phi|^2 R + \frac{1}{2} D_{\nu}\phi^{\ast}D^{\nu}\phi +  \frac{1}{2} D_{\nu}\Phi^{\ast}D^{\nu}\Phi - V_{bi}(\phi, \Phi) \, . \label{L_phi_Phi}
 \end{equation}
 
Similar locally scale invariant Lagrangians of the scalar field sector have been introduced and studied by several  author groups during the last few years.\footnote{\cite{Nishino/Rajpoot:2004,Meissner/Nicolai,Quiros:2014,Bars/Steinhardt/Turok:2014} and others. A similar form  of the scalar Lagrangian with {\em global scale invariance} is  considered in \cite{Garcia-Bellido/Shaposhnikov_ea:2012}. The last mentioned authors introduce their Lagrangian as the ``minimal scale invariant extension'' of the SM and GR. It could have been re-read without  in the sense of local scale invariance, had not other authors done so before. \label{fn Nishino ea}}
Of course,  the Lagrangians studied in these papers differ among each other.\footnote{All of the mentioned papers include  a direct coupling of the Higgs field to the scalar curvature, but conclude that the effects can be neglected. Some are fascinated by the perspective to study the role of the Higgs field for cosmological ``inflation''.  Meissner/Nicolai and Bars/Steinharst/Turok do not use a Weyl geometric framework but   consider conformal or ``Weyl scaling''. The last mentioned group of authors study the effects of considering $\phi$ as a ghost field (inverse signs of gravitational couplings of $\phi$ and $\Phi$ and inverse signs of kinematical terms) on geodesic completability of cosmological models. Although all investigations deserve attention in themselves we need not, and cannot, go into more  details here.}
 Here we consider (\ref{L_phi_Phi}) as a paradigmatic example and  concentrate on  the role of the gravitational coupling of $\phi$ for the emergence of a fixed (constant) value for $h(x)$ and, in this sense, for the ``spontaneous breaking'' of scale symmetry.

For that we have to investigate, whether a common ground state of the two scalar fields exist, that is, we have to ask for a (local) minimum of $V(\phi,h)$ in both variables. An easy calculation
 shows  that the gradient $grad\, V = (\partial_{\phi}V, \partial_h V )$ vanishes for $h_o = \alpha |\phi|_o, \; \phi_o^2 =-\frac{\xi^2}{\lambda'}R$, and that $V(\phi_o, h_o)$ is, indeed, a local minimum.\footnote{$\partial_{\phi}V= R\xi^2 \phi- \alpha^2\lambda \phi (h^2- \alpha^2 \phi^2)+ \phi^3 \lambda', \; \partial_h V = \lambda h (h^2 - \alpha^2 \phi^2) $, and for $\phi_o, h_o$ as above $Hessian\,(V)_{|(\phi_o,h_o)} >0$ (positive definite).  }
  That shows that  
 the ground state of the gravitational scalar field $\phi_o$, compared with (\ref{potential condition}),   is not affected by its coupling to $h$. The common ground state $(\phi_o,h_o)$ of the two fields is 
\begin{equation} \phi_o^2 =  -\frac{\xi^2}{\lambda'}R\,, \qquad  h_o^2 = \alpha^2 \phi_o^2 =  -\frac{\alpha^2 \xi^2}{\lambda'}R \, . \label{common ground state}
\end{equation} 

Like in section 4.4 we see that the {\em scalar field gauge} agrees with the {\em Weyl gauge}, if $\phi$ is in its ground state. Moreover, (\ref{common ground state}) shows that the ``Higgs field gauge'' (i.e., the gauge in which the Higgs field is scaled to a constant vacuum expectation value) is identical to scalar field gauge and to Weyl gauge: There is one gauge in which all three, gravitational scalar field, Higgs field, and Weyl geometric scalar curvature are scaled to a constant norm. By obvious reasons we call it {\em Einstein-Weyl gauge} and denote the respective values by $\phi_c, h_c, R_c$ (lower $c$ for ``constant'').

From empirical observation we get  constraints 
\begin{equation} \xi^2 \phi_c^2 = M_{pl}^2 \approx (2.4 \cdot 10^{18}\, GeV)^2\, , \qquad \alpha^2 \phi_c^2 = v^2 \approx (246 \, Gev)^2\, . \label{constraints}
\end{equation}
Therefore
\begin{equation} \frac{\xi}{\alpha} \sim 10^{16} \, , \quad \mbox{the ew-Planck hierarchy factor.}
\end{equation}  

If we assume $\lambda' \sim 1$,  we read off 
from (\ref{common ground state}) and (\ref{constraints})   that $\phi_c$ lies ``logarithmically in the middle''  between $R$ (in Einstein-Weyl gauge) and $M_{pl}$; i.e. $\phi_c$ is the geometrical mean between the two:\footnote{This relation may  lie at the bottom of some of the ``large number coincidences'' which fascinated Eddington,  to a lesser degree Weyl, and others.}
\[ |R|^{\frac{1}{2}}  \quad  \stackrel{\xi/\sqrt{\lambda'}}{ \longrightarrow} \quad |\phi_c| \stackrel{\xi}{ \longrightarrow} \quad M_{Pl}
\] 
For $R\, \dot{\sim} \, 10\, H^2$ with Hubble constant $H$ and $\hbar c H \sim 10^{-33}$ and $\lambda'\sim 1$, we find\footnote{For the estimates of $R$ and $H$ see footnote \ref{fn H in ev}.} that the order of magnitude of the second hierachy factor (between the energy level of the scalar field's ground state and Planck energy) is $\xi \sim 10^{30}$.  The ew scale lies  close to the geometrical mean between $\phi_c$ and $M_{pl}$:
\[ H  \quad  \stackrel{\xi}{ \longrightarrow} |\phi_c|  \quad  \stackrel{\alpha}{ \longrightarrow} \quad  v  \quad  \stackrel{\alpha'}{ \longrightarrow} \quad  M_{pl}\, ,
\] 
where $\xi = \alpha \alpha'$,  $\alpha \sim 10^{14}, \; \alpha' \sim 10^{16}$. The effective (classical) value of $\lambda$ is  constrained by the observational values of the Higgs mass $m_h\approx 126\, GeV$ and $v\approx 256\, GeV$ to $\lambda \approx 0.24$.\footnote{The tachyonic mass term of the Higgs field $\frac{\lambda}{4}\alpha^2 |\phi_c|^2 =\frac{\lambda}{2}v^2$ turns into a real mass term for the Higgs excitation, $m_h^2= \lambda v^2$, thus the value vor $\lambda$.} 

At first glance one might expect that $\lambda'$ is constrained by  dark energy considerations. But this is not the case, as one can  check by inspecting the changes in the energy tensor (\ref{Theta}) of the scalar sector after introducing the Higgs field.  In the ground state of the scalar fields the only changes arise from the contribution of the kinematical terms of $h$. They are suppressed like the ones of $\phi$ in the effective approximation (\ref{vacuum energy}, \ref{"dark matter"}).\footnote{ In (\ref{V(phi,h)}, \ref{L_phi_Phi})the Higgs field $\Phi$ is not coupled to $R$; therefore no boundary term of the variation of the Hilbert term appears. (This is similar for the direct Higgs coupling to $R$ considered by the authors mentioned in fn. \ref{fn Nishino ea}, because in all cases the Higgs coupling to scalar curvature is by far outweighed by the dominating $\phi$ term  ($\sim M_{pl}^2$)). The quartic term of $h$ does not deliver a  contribution to the energy tensor because in the ground state it is cancelled by the contribution from $\alpha^2 \phi^2$. The additional kinematical terms for $\Phi$, (those with factor $\xi^2$ in (\ref{Theta}) are suppressed as indicated in the main text.}
Thus the {\em energy  momentum tensor of the combined scalar fields $(\phi, \Phi)$ in their ground state is still given by (\ref{vacuum energy}), (\ref{"dark matter"})} like in the single gravitational scalar field case of section 4.4.

The often discussed question why the quartic term of the Higgs field does not dominate gravitational {\em vacuum energy} in the cosmological term finds  a completely convincing {\em  explanation on the classical level}.
Moreover,  while in Einstein gravity 
the  cosmological constant term results in the anomalous feature of vacuum energy of being able to influence the dynamics of matter and geometry without back-racting to them, this problematic feature is dissolved here (like in other JBD-like approaches).

 These are pleasing results of our investigation of the   intertwinement between the Higgs field and the gravitational scalar field. 
Let us resume the most important qualitative (structural) results:
\begin{itemize}
\item[$-$] The Higgs coupling to gravity  considered here does {\em not affect the energy-momentum tensor}  of the scalar sector.
	\item[$-$] In its ground state the intertwined two gravitational scalar fields adapt to the Weyl geometric scalar curvature like in the case of ``pure'' gravitational scalar field (section 4.4).
	\item[$-$] Therefore the vacuum energy not only influences matter and geometrical dynamics, but also back-reacts to the latter. 
\item[$-$] Different to what one finds in the respective literature,\footnote{Cf. fn \ref{fn Nishino ea}}
there is {\em no complete  decoupling} of the electroweak sector from gravity in the ``low'' energy regime \ldots, 
 \item[$-$] \ldots because the dimensional parameter $\mu^2$ of the ordinary Higgs mechanism is derived from the scale covariant coupling with the gravitational scalar field.
 \item[--] The ground state of the latter {\em is determined by the coupling to gravity} ($\xi^2 \phi^2 R$ term). 		
 \item[$-$] In this sense,  the  {\em two scalar fields}  are gravitationally combined like {\em twins}.\footnote{We may hope that a  deeper understanding of the emergence of the scalar field sector can lead to a common quantum field theoretical origin of the two related classical fields. }
  Only taken together they induce a kind of ``spontaneous breaking'' of (local) scale symmetry.
\end{itemize}
 The last point deserves to be discussed in more detail in the next section.

\subsubsection*{5.4 A Weylian hypothesis reconsidered}
\addcontentsline{toc}{subsection}{5.4 A Weylian hypothesis reconsidered}
The proportionality between the squared scalar field's value with $R$ has most important consequences for our understanding of measurement processes.
Quantum mechanics  teaches us how atomic  spectra depend on the mass of the electron.
% (\ref{electron mass}).\footnote{The  dimension-less coefficient $\mu_e$ of the electron must here, of course, include radiative corrections of the bare mass. } 
 The energy eigenvalues of the Balmer series in the hydrogen atom are governed by the Rydberg constant $R_{ryd}$,
\begin{equation} E_n =  - R_{ryd} \frac{1}{n^2} \, , \qquad  n \in \mathbb N \, . \label{Balmer}\end{equation}
The latter (expressed in electrostatic units) depends on  the fine structure constant $\alpha_f $ and on the electron mass,   thus finally on the norm of Higgs field:\footnote{Vacuum permissivity $\epsilon _o = (4\pi)^{-1}$; then $e^2=2 \alpha_f \epsilon _o h c = \alpha_f \hbar c$. }
\begin{equation} R_{ryd} = \frac{e^4 \, m_e}{2 \hbar^2}= \frac{\alpha_f ^2}{2} m_e c^2 = \frac{\alpha_f ^2}{2}  \sqrt{\mu _e} \, v c^2\,  \label{Rydberg constant I}
\end{equation}
This equation is a classical idealization; with  field quantization the fine structure constant $\alpha_f$, and with it $R_{ryd}$, become scale dependent.\footnote{C. H\"olbling and R. Harlander  made  me aware of this problem.}

In our scale covariant approach the  masses  of elementary fermions  depend on  indirect coupling to gravity as argued in 5.3. The Rydberg ``constant'' turns into a scale covariant quantity of weight $-1$ and  scales with $\phi$, while the electron charge is considered as a ``true'' (nonscaling) constant. In scalar  field gauge (in other words, in Einstein gauge) the Rydberg factor is also scaled to constant (on the classical level) together with $\phi$ and $h$. In terms of (\ref{common ground state}) it is 
 \begin{equation} R_{ryd} = \frac{\alpha_f ^2}{2}  \, \mu _e \, h_o\, c^2 =\frac{\alpha_f ^2}{2} \alpha \, \mu _e \, |\phi_o|\, c^2 \doteq \frac{\alpha_f ^2}{2} \alpha \, \mu _e \, |\phi_c|\, c^2\, .  \label{Rydberg constant II}
 \end{equation} 
Similarly, the usual atomic unit of length for a nucleus of charge number $Z$ is the Bohr radius $l_{Bohr} = \frac{\hbar}{Ze^2 m_e} $ and gets  rescaled just as well, like $|\phi|^{-1}$.

That is, typical {\em atomic time intervals} (``clocks'') and {\em atomic distances} (``rods'') are {\em regulated by the ew scalar field's ground state} $|h_o|$. If the discussion of section 5.3 hits the point, it is linked to the ground state of the gravitational scalar 
field and thus to Weyl geometric scalar curvature. Under the assumptions of section 5.3, 
a definition of units for central physical magnitudes like in the new SI rules establishes a measurement system in which {\em  the value of $|h|$  is set to a constant by convention},
 If it is evaluated in the framework of iWOD gravity (and presupposing the correctness of the  laws linking measurement procedures to natural constants on which  the SI regulations are based), that corresponds to fixing Einstein gauge for actual measurements.\footnote{Cf. fn. (\ref{fn SI}). Although the calculation of the spectral lines of  $^{133}$Caesium is  more involved,  the dependence on electron mass  remains.}

In the end, the scaling condition of Einstein gauge (= Weyl gauge) and (\ref{Rydberg constant II})  give a surprising  justification for an ad hoc assumption introduced by Weyl during his 1918 discussion with Einstein. Weyl conjectured that  atomic spectra, and with them rods and  clocks,  adjust  to the ``radius of the curvature of the world'' \cite[309]{Weyl:STM}. In his view, natural length units are  chosen in such a way that scalar curvature is scaled to a constant,  the defining condition of  what we  call {\em Weyl gauge}. 
In the fourth edition of {\em Raum - Zeit - Materie} (translated into English by H.L. Brose) he wrote:
\begin{quote}
In the same way, obviously, the length of a measuring rod is determined by adjustment; for it would be impossible to give to {\em this} rod at {\em this} point of the field any length, say two or three times as great as the one that it now has, in the way that I can prescribe its direction arbitrarily. The world-curvature makes it theoretically possible to determine a length by adjustment. In consequence of this constitution the rod assumes a length which has such and such a value in relation to the radius of curvature of the world. \cite[308f.]{Weyl:STM}
\end{quote}

The electroweak link  explored in section 5.3  thus underpins a feature of Weyl geometric gravity which was introduced  Weyl  in a kind of ``a priori'' speculative move. 
In the 5th (German) edition of {\em Raum - Zeit - Materie} Weyl already called upon   Bohr's  atom model as a first step  towards  justifying  his scaling conjecture:
\begin{quote}
Bohr's theory of the atom shows that the radii of the circular orbits of the electrons in the atom and the frequencies of the emitted light are determined by the constitution of the atom, by charge and mass of electron and the atomic nucleus, and Planck's action quantum.\footnote{``Die Bohrsche Atomtheorie zeigt, da\ss{} die Radien der Kreisbahnen, welche die Elektronen im Atom beschreiben und die Frequenzen des ausgesendeten Lichts sich unter Ber\"ucksichtigung der Konstitution des Atoms bestimmen aus dem Planckschen Wirkungsquantum, aus Ladung und Masse von Elektron und Atomkern \ldots'' \cite[298]{Weyl:RZM5}.}
\end{quote}

At the time when this was written, Bohr had already derived (\ref{Balmer})  and (\ref{Rydberg constant I}) for the Balmer series of the hydrogen atom and for the Rydberg constant \cite[201]{Pais:Inward}. Weyl saw, at first,   no reason to give up his scale gauge geometry. He  rather continued:
\begin{quote}
The most recent development in atomic physics has made it likely that the electron and the hydrogen nucleus are the fundamental constituents of all matter; all electrons have the same charge and mass, and the same is true for all hydrogen nuclei. From this it follows with all evidence that {\em the masses of atoms, periods of clocks and lengths of measuring rods are not preserved by some tendency of persistence; it rather is a result of some equilibrium state determined by the constitution of the structure (Gebilde), onto which it adjusts so to speak at every moment anew} (emphasis in original).\footnote{``Die neueste Entwicklung der Atomphysik hat es wahrscheinlich
gemacht, da\ss{}  die Urbestandteile aller Materie das Elektron und
der Wasserstoffkern sind; alle Elektronen haben die gleiche Ladung und
Masse, ebenso alle Wasserstoffkerne. Daraus geht mit aller Evidenz hervor,
da\ss{} {\em sich die Atommassen, Uhrperioden und Ma\ss{}stabl\"angen nicht durch
irgendeine Beharrungstendenz erhalten; sondern es handelt sich da um
einen durch die Konstitution des Gebildes bestimmten Gleichgewichtszustand,
auf den es sich sozusagen in jedem Augenblick neu einstellt}.'' (loc. cit., emph. in or.,  298)}
\end{quote}
The claim that ``it follows with all evidence'' was, of course, an overstatement. It is well known how  Weyl himself shifted his gauge concept from scale to phase only a few years later (in the years 1928-1929). After this shift he reinterpreted the Bohr frequency condition. In later discussions he referred to it as an argument {\em against} the physicality of his scale gauge idea.\footnote{Compare, for example, Weyl's  remarks in \cite[83]{Weyl:PMNEnglish}.}

This shift gives evidence to a paradoxical double face of Weyl's remarks with regard to the Bohr frequency condition.  For Weyl it may have contained a germ for the later distantiation from his first gauge theory, hidden behind an all too strong rhetoric of ``evidence''. But now  it appears again in a completely new light. 
Read in a systematical   perspective, Weyl's remarks from 1922/23 can  now even appear as foreshadowing  {\em  a  halfway marker  on the road towards  a  bridge between gravity and atomic physics}. Whether this bridge resists  depends, of course, on the  answer to the question whether or  not the link discussed here  between the scalar fields of gravity and  of ew theory  is  realistic (``physical''). This question is open for further research. 

At the end of the 1920s there was no chance for anticipating the electroweak pillar of the bridge. Historically, Weyl was completely right in considering the Bohr frequency condition as an indicator that his early scale gauge geometry could not be upheld  as a physical theory in its original form.
 Weyl's original interpretation of the scale connection as the electromagnetic potential  became obsolete  in the  1920s, but 
 his ad hoc hypothesis that {\em Weyl gauge} indicates measurements by material clocks and ``rods''  most directly may now get new support.\footnote{Cf. \cite{Scholz:2014Weylgauge}.}

\subsection*{6. Another look at cosmology}
\addcontentsline{toc}{section}{6. Another look at cosmology}
It is of   interest to see how cosmology looks from the vantage point of scale covariant gravity, not only in order to test  the latter's formal potentialities on this level of theory building but also because certain features of recent observational evidence of cosmology are quite surprising: dark matter and dark energy,  distribution and dynamics of dwarf galaxies,  lacking correlation of metallicity with redshift of galaxies and  in quasars (i.e, no or, at best, highly doubtful indications of evolution), too high metallicity in some deep redshift quasars and the intriguing, but as yet unexplained, distribution of quasar numbers over redshift.\footnote{\cite{Kroupa/Pawlowski:2010,Kroupa:2010,Sanders:DarkMatter,Hasinger/Komossa,%
Cui/Zhang,SDSS:2007,Tang/Zhang:quasars}. \label{fn anomalies}} 
 It would not be surprising if some of these develop into  veritable anomalies for the present standard model of cosmology. At least they  indicate that some basic changes in the conceptual framework for cosmological model building seems to be  due.

 At the moment we cannot claim that  these (potential) anomalies will be resolved by Weyl geometric gravity, neither are  cosmological investigations in the framework of Weyl geometry  bound to go beyond the general frame of the present  picture of an expanding universe plus ``inflation''. Often they ares still committed  to the latter.\footnote{E.g. \cite{Nishino/Rajpoot:2009,Quiros:2014}. }
But  the above mentioned problems  are sufficient reason for  reflecting  the status of present cosmology and to  compare it with  alternative approaches.    

Weyl geometric gravity is not  the only alternative ``on the market''; many others  are being  explored.\footnote{Some of them have been reviewed from a contemporary history view  in \cite{Kragh:varying_c,Kragh:multiverse,Kragh:oscillating} and the  (quasi) steady state approach  in \cite{Lepeltier:dialogue}. 
  Less discussed are different kinds of static or neo-static approaches \cite{Crawford:static_universe,Masreliez:I,Scholz:BerlinEnglish,Scholz:FoP}, or  explorations of unconventional views on vacuum energy like in \cite{Fahr:BindingEnergy}.} 
 The  number of publications  which accept the present standard cosmology in the observable part but develop alternatives to the ``big bang'' singularity seems to be rising.\footnote{Among them   \cite{Penrose:Cycles,Steinhardt:cyclic,Bars/Steinhardt/Turok:2014,Bojowald:Vor}.}
Some of them may be worth considering in  philosophical `meta'-reflections on cosmology, complementary to  philosophical investigations centered on   more mainstream  lines of investigation in  cosmology.\footnote{Very selectively, \cite{Smeenk:Inflation,Zinkernagel/Rugh:cosmic_time,Beisbart:cosmological_principle} and the recent volume {\em 46}  of {\em Studies in History and Philosophy of Science, Part A}.}
 
 \subsubsection*{6.1 Friedman-Robertson-Walker models  in iWOD gravity}
  \addcontentsline{toc}{subsection}{6.1 Friedman-Robertson-Walker models  in iWOD gravity}
%For cosmological models in our approach, it is an intriguing feature of scalar field gauge (\ref{scalar field gauge}) 
One often uses approximate descriptions of cosmological spacetime  by   models with maximal symmetric spacelike folia, i.e.,  {\em Friedman-Robertson-Walker (FRW)}  manifolds with  metric of the form
\begin{eqnarray} \tilde{g}: \quad   d\tilde{s}^2 &=&  d \tau^2 - a(\tau )^{2} d\sigma^2_{\kappa} \, ,  \label{warped metric} \\
 d\sigma^2_{\kappa} &=&  \frac{dr^2}{1-\kappa\, r^2} + r^2 ( d\Theta ^2 +   \sin^2\Theta \, d\phi^2 )\, . \nonumber
\end{eqnarray}
The underlying manifold is $M \approx I\times S^{(3)}$, with $I \subset \mathbb R$ and $ S^{(3)}$  three-dimensional.  $ S^{(3)}$  is endowed with a Riemannian structure of constant sectional curvature $\kappa$, locally parametrized by spherical coordinates $(r,\Theta,\phi)$.\footnote{Here  $\phi$ is the usual designation of an angle coordinate. Contextual reading  disentangles the dual meaning for $\phi$ we allow here. --- For a survey of  models with less symmetry constraints see \cite{Ellis/vanElst:Cosmological_Models}, but consider the argumentation in \cite{Beisbart:cosmological_principle}.}

For Weyl geometric FRW models the behaviour and calculation of cosmological redshift is very close to what is known from the standard approach. The energy  of a photon describing a null-geodesic $\gamma(\tau)$ considered by cosmological observers along trajectories of a  cosmological time flow unit vector field $X(p)$, $p\in M$, $X=x'(\tau)$,
 is  given   by $E(\tau)= g(\gamma'(\tau), 
X(\gamma (\tau)))$.\footnote{Cf. \cite[110, 116]{Carroll:Spacetime}, for Weyl geometric generalizations, e.g., \cite{Scholz:FoP,Poulis/Salim:2011,Romero_ea:2011}.} 
Cosmological redshift is expressed by the ratio 
\begin{equation}  z+1 = \frac{E(\tau_o)}{E_(\tau_1)} = \frac{g(\gamma'(\tau_o), X(\gamma (\tau_0)))}{g(\gamma'(\tau_1), X(\gamma (\tau_1)))} \, .
\end{equation} 
As we are working with geodesics of weight $-1$, $w(X)=-1$, and $w(g)=2$, energy expressions for photons with regard to cosmological observers are {\em independent} of scale gauge; so is {\em cosmological redshift}.

In the standard view the warp function $a(\tau)$ is  considered as an expansion of space with the cosmological time parameter
 $\tau$. After an embedding of Einstein gravity into  iWOD  this view is no longer mandatory.\footnote{Every Riemannian model  $(M, g)$ with Lorentzian spacetime $M$ and metric $g$ can easily be considered as an integrable Weyl geometric model with Weyl metric $[(g,0)]$. If the dynamics is enhanced by a scalar field and scalar curvature of the model is $\neq 0$ the  extension is dynamically non-trivial. For a discussion of  consequences for the view of gravitational effects see \cite{Romero_ea:2012}. }
Even more, it is no longer   convincing.  If  electroweak coupling -- or any other mechanism leading to an analogous scale gauge behaviour -- is realistic,  Friedman-Robertson-Walker geometries are better  considered in  Weyl gauge, i.e., scaled to constant scalar curvature in the Weylian generalization, than in Riemann gauge. In consequence, a large part of what appears as ``space expansion'' $a(\tau)$ in present cosmology, perhaps even all of it, is  encoded by the scale connection $\varphi$ after rescaling to Weyl gauge. 

In the result, the {\em cosmological redshift need not (exclusively) be due to expansion; it can just as well be a result of field theoretic effects expressed by the scale connection} (or, equivalently  in Riemann gauge, by a ``varying cosmological constant'' and  ``varying'' particle masses and measuring units, regulated by  the scalar field).\footnote{See  \cite{Scholz:BerlinEnglish,Scholz:FoP,Poulis/Salim:2011}.}
A similar argument that redshift may result from ``varying particle masses'' was recently given    in the framework of JBD gravity by  \cite{Wetterich:2013} .\footnote{Wetterich's reputation in the physics community helped to bring his argument into the  {\em Nature} online journal http://www.nature.com/news/cosmologist-claims-universe-may-not-be-expanding-1.13379.} 

The counter argument that a quantum mechanical explanation is lacking and a necessary prerequisite for accepting the explanation  is self-defeating, as the  explanation by space expansion does not provide one either. Expansion or scale connection, both are essentially (gravitational) field theoretic effects and, in a scale covariant theory, even mutually interchangable.

 \subsubsection*{6.2 A simple model class: Weyl universes}
 \addcontentsline{toc}{subsection}{6.2 A simple model class: Weyl universes}
 If we extend our view from the classical cosmological models built upon Einstein's theory to scale invariant gravity, the picture of the ``universe'' may change considerably. Models come into sight without any expansion at all,  where   the {\em whole cosmological redshift} is due to the scale connection $\varphi$. Toy models of such a type have been   studied in \cite{Scholz:FoP}.\footnote{The   balancing condition between matter and the scalar field assumed there did    not yet take the link to ew theory into account; therefore the dynamical assumptions of  \cite{Scholz:FoP} differ from those discussed here and lead only to provisional results.}
  The  constraint for the scalar field, established here by the potential condition (\ref{potential condition}), facilitates the analysis considerably  and allows to derive  a surprisingly simple uniqueness  result with regard to dynamic equilibrium.

In Riemann gauge, these models can be represented as particularly simple  Friedman-Robertson-Walker spacetimes with a varying scalar field (a ``varying gravitational constant'') and a  a linear warp (``expansion'') function   $a(\tau)= H \tau$.\footnote{Reparametrization of the time coordinate in Riemann gauge gives the picture of a ``scale expanding cosmos'' \cite{Masreliez:I} with exponential scale growth $ds^2 = e^{2HT}(ds^2 - d\sigma_{\kappa}^2)$. $H$ the Hubble parameter observed today, cf. fn (\ref{fn Hubble}). }
 Weyl gauge, on the other hand,  shows a  non-expanding spacetime,  of course now with a non-vanishing scale curvature which contains all the information of the former warp function. After reparametrization of the timelike parameter $\tau=H^{-1}e^{Ht}$,
 the Weylian metric is given by
\begin{eqnarray}    ds^2 &=& dt^2 - \left( \frac{dr^2}{1-\kappa\, r^2} + r^2 ( d\Theta ^2 +   \sin^2\Theta \, d\phi^2 ) \right) = dt^2 - d\sigma_{\kappa}^2 \,  \\  \varphi &=&(H,0,0,0) \, , \nonumber
\end{eqnarray} 
($d\sigma_{\kappa}^2$ the metrik on the spacelike folia of constant curvature). 
These models have been called {\em Weyl universes}, in particular {\em Einstein-Weyl }universes for $\kappa >0$  \cite{Scholz:FoP}. They are time homogeneous in a Weyl geometric sense.

The cosmological time flow remains static $x(\tau)= (\tau, \tilde{x})$ with $\tilde{x}\in S^{(3)}$. Coefficients of the Weylian affine connection are easily derived from the classical case, in particular $\Gamma^0_{00}=H $ and $\Gamma_{oi}^i=H\; (i=1,2,3) $ , while  all $\Gamma_{ij}^k$, for $i,j,k= 1,2,3$, are those of the spacetime folia (3-spaces of constant curvature). The parameter 
\begin{equation} \zeta:= \frac{\kappa }{H^2}
\end{equation}
characterizes Weyl universes up to  isomorphism (Weyl geometric isometries).

 The increment in cosmological redshift in Weyl universes is  constant, and thus
\begin{equation} z+1 = e^{Ht} \label{redshift}
\end{equation}
or $ z+1 = e^{Hc^{-1}d} $ for signals from a point of distance $d$ on $S^{(3)}$ from the observer (depending on ``which'' $H$ is meant, $H_o$ or $H_1$).\footnote{More precisely, one could distinguish between the time dimensional Hubble constant $H_o \approx 2.27\, 10^{-18}\, s^{-1}$ and its length dimensional version $H_1=H_0 c^{-1} \approx 7.57\, 10^{-29}\, cm^{-1} $ with its inverse,   the {\em Hubble distance} $H_1^{-1}\approx 4.28\, Mpc$.  \label{fn Hubble} }
In Weyl gauge it is described by the time component of the scale connection, $\varphi_o=H$. 

Ricci  curvature (independent of scale gauge) and scalar curvature in Weyl gauge are\footnote{Cf., e.g., \cite{ONeill}, or any other textbook about Robertson-Walker spacetimes.}
\begin{eqnarray} Ric&=&  2 (\kappa +H^2)d \sigma_{\kappa }^2\, ,   \label{Ric Weyl universes} \\
R &=& - 6 (\kappa +H^2) \, . \label{R Weyl universes}
\end{eqnarray}
In Weyl gauge the left hand side of the generalized Einstein equation (\ref{massless Einstein equation})  has timelike component $3(\kappa +H^2)$ and spacelike entries $(\kappa +H^2)g_{ii}$, i.e.  $ -(\kappa +H^2)d\sigma_{\kappa  }^2$,  ($i= 1, \ldots , 3$). That is familiar from classical static universe models. The absolute value of negative pressure $p \,g_{ii}$ is here  $|p|= \kappa +H^2$, i.e., one third of the energy density $3(\kappa +H^2)$. The  only difference to classical Einstein universes is marked by  the $H^2$ terms.

In Einstein gravity, static universes are stricken by tremendous problems, even inconsistencies, with regard to  their dynamics. It turned out impossible to stabilize them by a cosmological vacuum energy term or  by substitutes. That is  different for the  energy momentum of the scalar field. 
Calculation of the scale covariant derivatives of $|\phi|^2 $ fror Weyl universes leads to\footnote{Note that the scale covariant derivative of a function $f$ of weight $w(f)=-2$ need not be zero, even if  $f$ is gauged to a constant. For Weyl universes  $D_0 f = - 2 H f$ and $D_0D_0f =D_0 D^0 f = 6 H^2 f$, because of $\Gamma_{oo}^o=H$. Moreover, $D_1D^1 f = -2 H^2 f$,  similarly for $j=2,3$ because of $\Gamma_{oi}^i = H$ for $i, j = 1, 2, 3$; thus $D_{\nu}D^{\nu}|\phi|^2=0$
(wrong calculation in \cite{Scholz:FoP}, corrected in \cite[64]{Scholz:AdP}).} 
\begin{eqnarray} 
\Theta^{(I)} &=& \frac{3}{2}(\kappa +H^2)g \, \\
\Theta^{(II)} &=& \mbox{diag}\,(6 H^2 g_{00}, -2H^2 g_{11}, -2H^2 g_{22},-2H^2 g_{33})\, 
\end{eqnarray}  
(\ref{vacuum energy}),   (\ref{"dark matter"}), (\ref{vacuum Theta I}). Comparison with (\ref{Ric Weyl universes}, \ref{R Weyl universes}) shows that  the Einstein equation  holds for exactly one value of the spatial curvature,\footnote{$\kappa =3\, H^2$ corresponds to $\Lambda = 6H^2$ with relative value $\Omega_{\Lambda } =2$. Note that the ``dark matter''  term $\Theta^{(II)}$ has  positive pressure, characterized by $\frac{p}{\rho}= \frac{1}{3} $, and contributes  $\Omega _{\Theta^{(II)}} = 2$ to the relative energy density.}
 \begin{equation} \kappa_o = 3\,H^2 \, ,  \; \mbox{i.e.  }\; \zeta_o =3 \, , \quad \mbox{then}\; \; R_o =-24\,H^2.\label{stable Einstein universe}
 \end{equation} 
A  heuristic consideration indicates that   the Einstein-Weyl model  with $\zeta =3$ seems to be {\em  stable} inside the parameter space of Robertson-Walker spacetimes without matter.\footnote{If space curvature varies  (under the constraint of constant spacelike curvature)    to $\kappa = \kappa _o+ \Delta $, both  the energy density $\rho$ {\em and} the absolute value $p$ of the negative pressure of the scalar field increase  by $\frac{3}{2}\Delta  $. The equilibrium condition known from the classical case  requires $\rho= 3p$ (Raychaudhury equation in the simplest case).  For $\Delta > 0$, i.e., comparatively ``too small'' radius of curvature, the negative pressure wins over contractive energy density of the scalar field and spacelike geometry  expands; for $\Delta <0 $ the dynamics works the other way round. This  indicates that the {\em scalar field} of iWOD gravity   {\em pushes spacetime on large scales towards an Einstein-Weyl universe} with parameter  $\zeta_o = 3$ and stabilizes it there. This heuristic consideration is supported by numerical simulations.}

We shall call this special case the {\em balanced Einstein-Weyl universe}.  Of course, more detailed investigations of the dynamical behaviour are  necessary.    It would be particularly  interesting to see whether the Einstein-Weyl universe,  $\zeta_o = 3$, is stable even under weaker symmetry conditions, perhaps even without any.  That will be difficult to investigate; but if so, it would give strong theoretical support for this model.\footnote{There seem to be certain analogies to  Hamilton flow in the study of the Poincar\'e  conjecture. One might conjecture that the scalar field evolves the spatial folia toward the maximally symmetric case.}

The stabilization of the Einstein-Weyl universe needs no additional matter besides the energy momentum contribution of the scalar field. In fact, the relative value of energy density of the scalar field, compared  with the critical density $\rho _{crit}=\frac{3H^2}{8\pi G}$,  is here $\Omega_{\phi}:= \Theta_{00}/(3H^2)= 4$. The contribution of the vacuum energy component is $\Omega_{\Lambda}=2$, supplemented  by the   same amount of  the ``dark matter'' like component $\Theta^{(II)}$. 

The present estimate for  baryonic matter, $\Omega_{bar}\approx 0.04$, is just one percent of it.  Inside a balanced Einstein-Weyl universe, such a tiny amount of baryonic matter could impress only small perturbations onto the symmetric spacetime solution, even if it is highly inhomogenously distributed. Then there seems to be ample space for distracting parts of the energy (and pressure) content of $\Theta^{(II)}$, making it slightly more inhomogeneous than it appears  in our idealized, completely homogeneous, vacuum case. Parts of it could easily deliver the dark matter effects detectable by dynamical deviation from local Newtonian mechanics (galaxy rotation curves) or by gravitational lensing. 

Without doubt, the surprisingly high value for dark energy $\Omega_{\Lambda}$ seems to indicate that our model is too far away from observational cosmological evidence to be taken seriously. But we have to pose the question how stable present precision values of cosmological observables are against shifts in the background theory on which the evaluation of empirical raw data relies.
 
  \subsubsection*{6.3 Theory ladenness of cosmological observations}
   \addcontentsline{toc}{subsection}{6.3 Theory ladenness of cosmological observations}
Positive curvature for spatial folia and  static geometry stand in harsh contrast to many features of the present standard model of cosmology. Moreover, observational evidence of the cosmic microwave background CMB and from supernovae magnitude-luminosity characteristics, measured with such impressing precision during the last decades, seem to outrule balanced Einstein-Weyl universes. At  first glance all that seems to speak against our simple model.

But we should be careful. If we want to judge the empirical reliability of a new theoretical approach we have   to avoid  rash claims of refutation on the basis of empirical results which have been evaluated and interpreted in a theoretical framework  differing in basic respects from the new one. {\em Theory-ladenness of the  interpretation} of empirical data is {\em particularly strong in the realm of cosmology}.  Enlarging the symmetry of the Lagrangian by scale invariance  comes down to a {\em drastic shift in the  constitutive framework} for the formulation of physical laws. Judgement of such a shift demands  careful comparative considerations. That has to be kept in mind in particular for the evaluation and conclusions drawn from the high precision studies of the cosmic microwave background (Planck and WMAP data).

In the Weyl geometric approach, cosmological redshift looks  like a field theoretic effect on the classical level; it is modelled by the (integrable) scale connection rather than by ``space expansion''.  The CMB seems to be just as well explainable  be a quantum physical background equilibrium state of the Maxwell field excited by stellar and quasar radiation,  as by the relic 
radiation of the standard picture.\footnote{Already I.E. Segal argued that on  an Einstein universe the quantized Maxwell field will, under very general assumptions, build up an equilibrium radiation of perfect Planck characteristic \cite{Segal:CMB1}.} 
The correlation of the tiny inhomogeneities in the temperature  distribution with large scale matter structures would  be independent of the causal evolution  postulated in the present  structure formation theory. It has to be checked whether the  flatness conclusion from CMB data is  stable against a corresponding paradigm change.
 
Supernovae data have to be reconsidered in the new framework, in particular with view on possible observation selection effects.\footnote{For a  detailed argument that strong observation selection effects may come into the play in the selection procedures of the SNIa data see \cite[sec. 4.6]{Crawford:static_universe}, for a first glance at supernovae data from the point of view of Einstein-Weyl universes  \cite{Scholz:FoP}.} 
Galaxy evolution would look completely different, as no big bang origin would shape the overall picture. In particular Seyfert galaxies and quasars can be understood as {\em late} developmental stages of mass accretion in massive galactic cores. Jets emitted from them seem to redistribute matter recycled after high energy cracking inside galactic cores. Structure  formation would have to be reconsidered.\footnote{For a sketch of such a picture see   \cite{Crawford:static_universe} or \cite{Fischer:universe}.} 
Nuclear synthesis would no longer appear as ``primordial'' but could take place  in stars on a much larger time scale than in the recieved view, and in galactic cores, respectively quasars. Then the {\em Lithium 6/7 riddle} might dissolve quite  unspectacularly.

Regenerative cycles of matter mediated by galactic  cores, quasars and their jets are excluded as long as  cosmology is based on Einstein gravity  by  the extraordinary role of its singularity structures (``black holes''). But these have to be reconsidered in the Weyl gravity approach. 

Because of the  Weyl gauge condition,  local clocks tick slower   in regions of strong gravity (large $_g{\hspace{-0.1em}}R$)  also in comparison with Riemann gauge. The resulting conformal rescaling demanded by the potential condition (\ref{potential condition}), Weyl gauge as Einstein gauge, and their influence on the rate of spectral clocks  (\ref{Balmer}) changes the picture of the spacetime metric near singularities of the Riemann gauge (and also in comparison to Einstein gravity). We cannot be sure that the singularity structure is upheld.   Conformal rescaling may change the whole geometry, similar to the effect that an  initial singularity may be due to a ``wrong'' (Riemannian) scaling of Friedman- Robertson-Walker geometry in the case of Einstein-Weyl universes. Such investigations have started for Weyl geometric gravity by \cite{Prester:2013} and in a different perspective by \cite{Bars/Steinhardt/Turok:2014}.

Much has to be done. But why should one head toward such an enterprise of basic reconsideration of the cosmological overall picture? Only a few astronomers or astrophysicists dare to tackle this task at the moment. Among them,  David Crawford has been investigating for  some time, how well different classes of observational  evidence fits into the picture of a comological model with static spherical   spatial folia. The outcome is not disappointing for this assumption  \cite{Crawford:static_universe}. The choice between an expanding space model or a (neo-)static one seems to be essentially determined by underlying (explicit or implicit) principles of gravity theory.\footnote{Crawford assumes a peculiar dynamics of ``curvature cosmology'' which claims to remain in the framework of Einstein gravity. It seems doubtful that this conception can be defended. But here we are mainly interested in the detailed investigation of observational evidence in parts I, II of \cite{Crawford:static_universe}.}

Certain basic problems of the the standard picture are   being discussed in the present discourse on cosmology. There are  different strategies to overcome them. The most widely  known  approaches for explaining the unexpected outer galaxy dynamics ascribe these effects  to ``dark matter'' \cite{Sanders:DarkMatter}.  
On larger scales the  evolution and distribution of quasars deliver already plenty empirical evidence, not so well in agreement with the ``old'' picture. Quasar data of the Sloan Digital Sky Survey (SDSS), the 2dF group,  and others outweigh  the supernovae observations in number, precision and redshift 
range \cite{Tang/Zhang:quasars,SDSS:2007}. A striking feature is that there is {\em no indication of evolution of metallicity} in quasars or galaxies along the cosmological timeline, i.e., in correlation to 
redshift.\footnote{Another, at the moment isolated, inconsistency with the received picture of  metallicity development is a quasar with redshift $z\approx 3.91$ and of  extremely high metallicity (Fe/O ratio about 3) observed by   \cite{Hasinger/Komossa}. Still it is  considerd as irritating only for the standard picture of star, galaxy, and quasar evolution \cite{Cui/Zhang}. But it could foreshadow more.}
Less well known, but perhaps even more important, are recent observations of distribution and dynamics of dwarf galaxies. They seem to indicate a fundamental inconsistency with the structure formation theory of the standard approach \cite{Kroupa:2010}.

Such irritating observations, combined with diverging research strategies, are  a worthwhile object for  metatheoretical investigations in a pragmatic sense. 
The concentration on new classes of observational evidence is often crucial for the process of clarifying mutual vices and virtues of competing theories.  That is the reason why we want to have a short glance at quasar distribution before we finish. 

  \subsubsection*{6.4  A geometrical explanation of quasar distribution?}
  \addcontentsline{toc}{subsection}{6.4  A geometrical explanation of quasar distribution?}
The {\em distribution of quasars} in dependence of redshift  shows a  distinctive  asymmetric bell shape with a soft peak  between $z \approx 0.9$ and $1.6$ and at first a rapid, then slackening, decrease  after $z \approx 2$ shown in figure 1.\footnote{Best data come from the 2dF collaboration and the Sloan Digital Sky Survey \cite{SDSS:2007,Tang/Zhang:quasars}. Here we take the data of SDSS 5th data release; total number of objects 77429 (fig. 1 upper curve), SDSS  corrections for selection effects reduces the total number by half \cite{SDSS:2007}; the total number of the corrected collective is 35892. The maximum of the corrected distribution is manifestly a little above $z\approx 1$; the authors give $z=1.48$ as the median of the collection. \label{fn SDSS}} 
%
%\newpage
%\thispagestyle{empty}
\begin{figure}
\vspace*{-10em}
\center{\hspace*{-10em} \includegraphics*[scale=1]{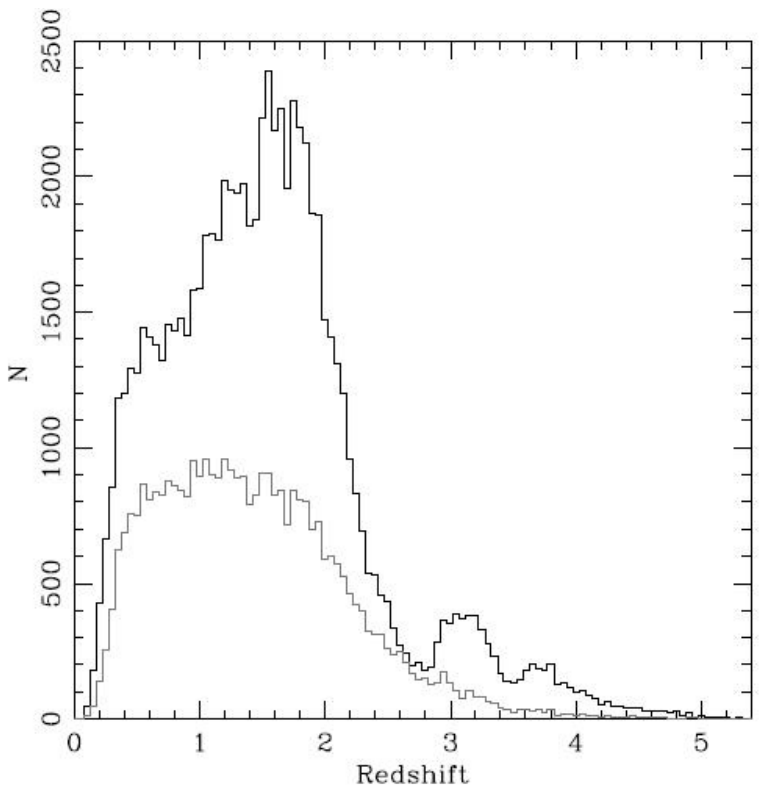}}
%\center{\includegraphics*[scale=0.7]{SDSS2007fig3}}
\vspace*{-20em}
\caption{ 
Redshift distribution of quasars from SDSS, 5th data release, width of redshift bins 0.05; upper curve raw data, lower curve corrected for selection effects; source (Schneider e.a. 2007, Fig. 3).}
\end{figure}
 \begin{figure}
\center{\includegraphics*[scale=0.75]{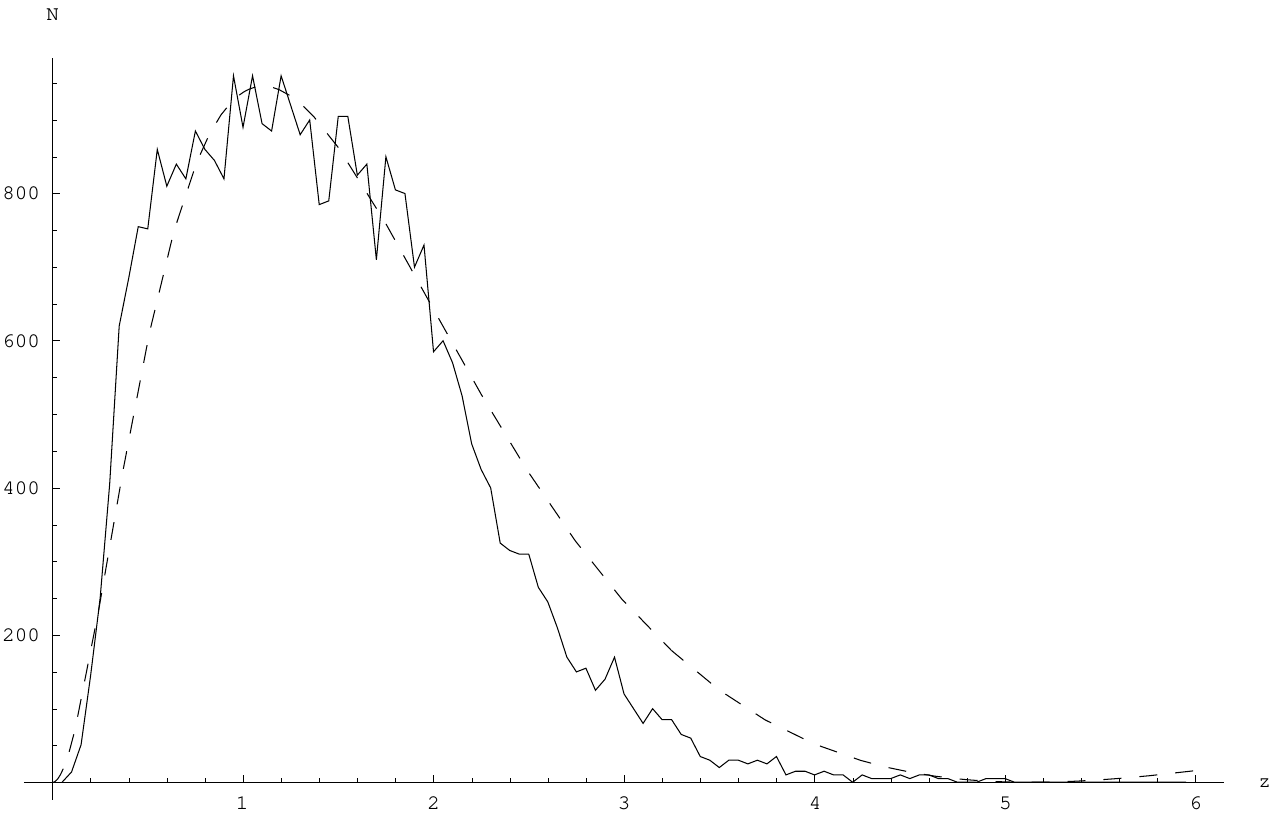}}
%\center{\includegraphics*[scale=0.75]{quasarverteilungA}}
\vspace*{-15em}
\caption{Redshift distribution of quasars from SDSS, 5th data release, corrected for selection effects (zig-zag curve), in comparison with equally distributed objects, volume increments over redshift bins of width 0.05, in Einstein-Weyl universe $\zeta = 3$ (dotted curve).}
\end{figure}
 In standard cosmology the regular distribution curve is a riddle which calls for ad hoc explanations of quasar formative factors.  From our point of view, the distribution pattern would be easy to explain: It turns out to be {\em close to the  volume increments of the backward lightcone} with rising redshift in the balanced Einstein-Weyl universe (fig. 2).\footnote{The maximum is reached around the equator of the spatial sphere. For $\kappa =3H^2$ the equator corresponds to redshift $z_{eq}=e^{H \frac{\pi}{2}(\sqrt{3}H)^{-1} }-1 \approx 1.47$ (\ref{redshift}). }

The deviation of the SDSS number counts from the calculated curve of the balanced 
Einstein-Weyl universe consists  of  fluctuations and  some remaining, rather plausible, observational selection effects: a moderate excess of counts below $Z=1$ and a suppression of observed quasars above $z \approx 2$. All in all, {\em the curves agree surprisingly 
well} with the assumption of  an {\em  equal volume distribution of quasars in large  averages} in the stable Weyl universe.

But there arises a  new question: 
The conjugate point on the spatial sphere is  reached at $z=e^{H \pi /c}-1=e^{\frac{\pi}{ \sqrt{3} } }-1 \approx 5.13$ ($r=\frac{1}{\sqrt{\kappa} }$ radius of the sphere). Interpreted in this model,  quasars and galaxies with higher redshift than $5.13$ ought to be images of objects ``behind'' the conjugate point and should  have counterparts with lower redshift on ``this'' side. For  terrestrial observers the two images are antipodal, up to the influence of gravitational deflection of the sight rays.  In principle, it should be possible  to check  the ``prediction'' of the Einstein-Weyl model of {\em paired antipodal objects} for the highest redshift quasars and galaxies  with present observational techniques.\footnote{The pairing of redshift and magnitudes are easy to calculate. But gravitational deflection of light disturbs the direction and local deviation from spherical symmetry close to the conjugate point blurs the focussing of light rays and affects magnitudes and redshift. Therefore an effective  decision of this question could be a true challenge for observational cosmology.}

At the moment such consequences have not yet been studied in sufficient detail. Maybe they  never will, unless some  curiosity of experts in gravity theory and in cosmology, both theoretical and observational, is directed  towards studying some of the more technical  properties of the iWOD approach.

For the `metatheoretical' point of view, it becomes apparent already here and now, that important features of  our present standard model of cosmology are not as firmly anchored in empirical evidence as is  often  claimed.  They are highly dependent on the interpretive framework of Riemannian geometry which plays a  constitutive role for 
Einstein gravity. Although we have very good reasons to trust this framework on  closer, surveyable cosmic scales (at least on the solar system level), it is not at all  clear whether we ought to trust its extrapolation to the gigantic scales  far above cluster level. The proposal of modified Newtonian dynamics (MOND) for explaining galaxy rotation curves  may be a sign that we  cannot   be sure  that  Einstein gravity describes gravity with the necessary precsision  already at outer galaxy level.\footnote{For other anomalous evidence see fn. \ref{fn anomalies} and, in particular,  the above mentioned study of dwarf galaxies in \cite{Kroupa:2010}.}

\subsection*{7. Review of  `transitions'}
\addcontentsline{toc}{section}{7. Review of  `transitions'}
We have seen how Weyl geometry offers a well structured intermediate step between the  conformal and the projective path structures of physics and  a fully metrical geometry (section 2). Riemannian geometry is only slightly generalized, if the Weyl geometric scale connection is integrable. Quantum physics gives convincing arguments to accept this constraint for considerations far below the Planck scale (Audretsch/G\"ahler/Straumann, section 2.3, and mass of the ``Weyl boson'',  section 4.2). As the Lagrangian of elementary particle physics is (nearly) invariant under point-dependent rescaling, a scale invariant generalization of Einstein gravity is a natural, perhaps necessary, intermediate step for  bridging the gap between gravitation theory and elementary particle fields. There are encouraging indications that integrable Weyl geometry  may  be of  help for the search of  deeper interconnections between gravity and quantum structures. Recently,  \cite{Codello_ea:2013} have proposed a quantization procedure of a Weyl (scale) invariant classical Lagrangian, which preserves Weyl invariance for the effective (quantized) action. 
Some experts expect a resolution of the notorious fine tunig problem for the Higgs mass from such a  move.\footnote{In a scale invariant Lagrangian the radiative corrections to the Higgs mass are expected to become only logarithmic rather than quadratic \cite{Bars/Steinhardt/Turok:2014}; for global scale invariance see the similar argument in \cite{Shaposhnikov/Zenhaeusern:2009}. }

In the 1980s Jordan-Brans-Dicke theory was explored for similar reasons, although in a different theoretical outlook and, up to now, without  striking success \cite{Kaiser:mass,Kaiser:colliding}. 
 A conceptual look at  Jordan-Brans-Dicke theory shows that  the latter's basic assumptions presuppose, usually without being noticed, the basic structure of integrable Weyl geometry (section 3). From a metatheoretical standpoint it seems surprising that this has  been  acknowledged explicitly  only very recently.\footnote{See  the first preprint version of this paper,  arXiv:1206.1559v1 and  \cite{Quiros_ea:2012} which was first posted on arXiv in 2011.}
  The Weyl geometric view   makes some of the underlying assumptions clearer and  supports the arguments of those who  consider the Einstein frame as the ``physical'' one (although this is an oblique way of posing the question). Physicists often seem to  withhold from such   metatheoretical considerations by declaring them as formal -- and ``thus'' --   idle games. Philosophers of physics are of a different opinion. That this game  is not idle at all, can  be seen  by looking at the transition from JBD theory to Omote-Utiyama-Dirac gravity (WOD). WOD  gravity has a Lagrangian close to JBD theory, but is explicitly formulated in   Weyl geometric terms (section 4). Historically, the transition from JBD to WOD gravity took place in the 1970s; but only a tiny minority of theoreticians in gravity and field theory contributed to it from the 1980s and 1990s until the present.\footnote{Of course other contributions  could be mentioned. Perhaps most extensive,  and not yet mentioned here, are the contributions of N. Rosen and M. Israelit, 
 cf. the provisional survey in \cite{Scholz:Mainz}.}

Perhaps the  mass factor of the scale connection (``Weyl field'') close to  Planck scale contributed to  the widely held  belief  that Weyl geometric gravity is  an empty generalization as far as physics is concerned.   We have shown that this is  not  the case. Although the scale connection $\varphi$ is able to play the role of a dynamical  field  only close to the Planck scale  -- where it may be important for a transition to quantum gravity structures -- it is an {\em important geometric device} for studying the dynamics of the interplay of the Weyl geometric scalar field with measuring standards (scale gauges) on lower energy scales. It is therefore not negligible even  in the integrable version of Weyl-Omote gravity  and closely related to  the scalar field $\phi$ which has to be considered as the {\em new dynamical entity} in the integrable case. The latter  may  represent  a state function of a quantum collective close to the Planck scale.

 By conceptual reasons iWOD does not need  breaking of scale co- or invariance; it allows to introduce scale invariant observable magnitudes with reference to any scale gauge of the scalar field (section 4.5). There are  physical reasons, however, to assume  such  ``breaking'' of a spontaneous type, if one takes the potential condition for the scalar field's ground state into account. A quartic potential of Mexican hat type arises here from the gravitational coupling of the scalar field. Formally, it is so close to the potential condition of the Higgs mechanism in electroweak theory that it invites us  to consider an extension of the Weyl geometric scalar field to the electroweak sector (section 5). We then recover basic features of the so-called Higgs mechanism of electroweak theory, but now without assuming an elementary field with an `ordinary' mass factor in the classical Lagrangian. From a metatheoretical point of view this closeness allows to illucidate the usual narrative of "symmetry breaking'' in the electroweak regime. We have shown how the {\em mass acquirement of weak bosons and elementary fermions comes about by a bridge between the Higgs field to gravity} via a coupling of the two  scalar fields (section 5.3). But of course we cannot judge, at the moment, whether such a link indicated by iWOD is more than a seducive song of the syrenes. 

From the point of view of the iWOD generalization of Einstein gravity we have reasons to seriously  reconsider our view of cosmology. The potential condition established by the electroweak link of the scalar field  ``breaks''  scale symmetry most naturally  in such a way that Weyl geometric scalar curvature is  set to a constant.  That corresponds to an idea of Weyl formulated in 1918 (section 5.4). It forces us to have a new look at  the Friedman-Robertson-Walker models of classical cosmology, re-adapted to the Weyl geometric context. 

The consequences of such a shift  cannot yet be spelled out in  detail. Models of constant scalar curvature and time homogeneity (Weyl universes) show interesting unexpected features.  The Einstein-Weyl universe with $\kappa =3H^2$ becomes a {\em dynamically consistent vacuum solution}. It seems to be  stabilized by the scalar field's energy momentum (section 6.2). Certain empirical data, in particular from quasar distribution and from  metallicity, speak against outright dismissing this model as counterfactual (section 6.4). 

In this framework, {\em dark energy } changes its character already at the classical level. It is generated  by the metric proportional part of the energy momentum of the scalar field $\Theta^{(I)}$.   Not only does it influences spacetime geometry, but it also reacts  back to curvature. 
In addition, the question of {\em dark matter} might get a new face, if the respective gravitational effects  can be explained by    the   part of the scalar field's energy momentum, $\Theta^{(II)}$, not proportional to the metric.   At the moment this is only a speculation; an important open question would be to study   the quantitative behaviour of inhomogeneities of $\Theta^{(II)}$ around galaxies and clusters in the iWOD approach.

In the end, the question is whether a MOND-like phenomenology can be recovered for constellations modelling galaxies by Weyl geoemtric  gravity. At the moment it seems that the static non-homogeneous isotropic vacuum solutions of iWOD  reduce to the  Schwarzschild-deSitter family of Einstein gravity with constant scalar curvature ($\neq 0$). If a Birkhoff-type theorem holds in iWOD gravity, it would be the only one. A chance for recovering MOND phenomenology may lie in the study of a  modified gradient term $L_{\phi}$ of the scalar field, similar to the one of Bekenstein/Milgrom's AQUAL theory. A first look at an  adaptation of this approach to the Weyl geometric setting is encouraging, at least from a conceptual point of view, maybe also from the point of dynamics; but that has to be judged by the experts in the field.\footnote{\cite{Scholz:MOND-like}}

Finally there is a fundamental argument in favour of the  model. A (neo-) static universe of the Einstein-Weyl type {\em could  bring back  energy and momentum conservation to  cosmology}. The ``expanding'' universe with its permanent increase of energy in the observed part of the universe by the cosmological constant term (``dark energy'') has very unpleasant consequences  for the asymptotics of local field constellations. 
  Einstein-Weyl universes have a group of automorphisms of type $SO(4)\times \mathbb R$, inside the larger group of (``gauge like'') diffeomorphisms as in  Einstein general relativity. For local inhomogeneities constellations with  Einstein-Weyl asymptotics, we may therefore expect that asymptotic time homogeneity symmetry $( \mathbb R, +)$ and the 6 spacelike symmetry generators of the cosmological model  lead to integral charge conservation for (on-shell) field constellations.\footnote{Asymptotically ``conserved'' (i.e. closed) $(n-2)$-forms derived from the superpotentials of Noether II currents have, in many similar cases,  been shown to lead to conserved charges defined by the flux of the superpotential forms through the asymptotic (closed)  $(n-2)$-dimensional boundaries  of spacelike hypersurfaces \cite{Weyl:1929,Abbot/Deser:1982,Barbich/Brandt:2002}. See A. Sus' contribution to this volume.  } 
Already this  difference to the expanding space view might invite physicists and philosophers alike to seriously consider the advantages  of a paradigm shift from the expanding view to the Einstein-Weyl framework, even though many of the deeply entrenched convictions of present cosmology had to be given up.

If course we had to give up   the received view of cosmological redshift as an effect of ``space expansion'' and substitute it by    an effect of the Weylian scale connection (section 6.1). Rescaling of the metric, in particular in regions of strong gravity (high Riemannian component of scalar curvature), changes the effective measure of time and length so strongly that in this regime no immediate transfer of geometrical results derived in classical gravity to the new context is possible. It is  no longer clear that  cosmological geometry  necessarily contains an initial singularity, nor even  localized singularities. Their external dynamics might be caused by finite matter concentrations which mimick structures of the black hole type if considered in Einstein gravity. 

Let us, at the end, come back to the philosopher quoted at the beginning of this article. 
 Herbart -- talking about metaphysics -- described   transitions between established theories, which he called the ``different formative stages'' of knowledge, as {\em revolutions}  which have to be traversed  before   research can  generate  concepts  necessary for  a ``distinguished enduring'' state 
\cite[198, 199]{Herbart:1825}. Also he  spoke of  the ``manifold delusions (mannigfaltige T\"auschungen)''  which our knowledge has to pass  before such an  enduring state can be reached.
 Riemann considered these remarks   important enough for  excerpting them.\footnote{``Wie die astronomische Betrachtung, die in die Tiefen des Weltbaues hinausgeht, so mu\ss{} auch die metaphysische Forschung, welche in die Tiefen der Natur eindringt, mancherley Revolutionen durchlaufen, ehe sie so gl\"ucklich ist solche Begriffe zu {\em erzeugen}, welche der Erscheinung genugthun und mit sich selbst zusammenstimmen'' \cite[198, emph. in original]{Herbart:1825}. The section ends by the remark ``Daraus folgt dann sogleich, {\em da\ss{} auch die T\"auschungen, die in diesem Werden nach einander entstehen, sehr mannigfaltig, da\ss{} sie den verschiedenen Bildungsstufen angemessen sind, welche successiv erreicht werden;} \ldots (ibid, 199). For Riemann's Herbart studies see \cite{Scholz:Herbart}.}

It seems that, also in cosmology, we   have to leave behind  ``manifold delusions'', before we have a chance to arrive at an   enduring picture (if at all) of how the universe in the large and the foundations of physics may go together.\\[4em]

\subsubsection*{Acknowledgements} 
\addcontentsline{toc}{section}{Acknowledgements} 
I thank G. Ellis  for his interest in the Weyl geometric approach to gravity  and for his comments. Thanks also go to the  referees, in particular to P.  Mannheim; their detailed remarks  helped  to improve content, structure and readability of the paper. M. Kr\"amer, R. Harlander and C. H\"olbling gave advice with regard to section 5.3. F. Hehl was so kind to give helpful comments in spite of his general scepticism with regard to the integrable Weyl geometric approach to gravity  (for his critical remarks see his contribution to this volume).  Most of all I want to express my gratitude to Dennis Lehmkuhl, the main editor of this book. Without his interest   and the  discussions in our interdisciplinary group on {\em Epistemology of the LHC}, based at Wuppertal,  this essay would not have been written.

\subsubsection*{Postscript}
The first version of this paper was written in summer 2012, some months  before the the Higgs detection was announced. Until the  publication, more than two years went by. This  gave plenty occasion for  rethinking basic questions of Weyl geometric gravity. Clear evidence of the author's  ``manifold delusions'' is  documented in the successive versions of this paper in arXiv:1206.1559. \hfill \\[0.1em]
{\em E.S., January 2015.}
\\[1em]
\newpage
%\newpage
%\nocite{Trautman:EPSa}
\footnotesize
\addcontentsline{toc}{section}{References}

 %\bibliographystyle{apsr}
 % \bibliography{a_lit_hist,a_lit_mathsci}

\end{document}